\documentclass[twocolumn, a4paper]{article}

\usepackage{nolta2026}
\usepackage{bm}
\usepackage{amsfonts}
\usepackage{amsmath,amssymb}
\usepackage{physics}
\usepackage{color}
\usepackage{txfonts}
\usepackage{graphicx}
\usepackage{subcaption}
\usepackage[sorting=none, style=numeric-comp]{biblatex}
\begin{document}

\title{Spatio-temporal structures in frog chorus with two species \\examined by laboratory experiments and mathematical modeling}

\author{
  Kanato Kawaguchi${}^\dag$, Ryu Takeda{}$^\star$, and Ikkyu Aihara{}$^\ddag$}

\address{
$\dag$ Graduate School of Systems and Information Engineering, University of Tsukuba, \\
Ibaraki 305--8573, Japan \\
$\star$ SANKEN, University of Osaka, Ibaraki, Osaka, 567--0047, Japan, \\
$\ddag$ Division of Information Engineering, Institute of Engineering, Information and Systems, \\
University of Tsukuba, Ibaraki 305--8573, Japan \\[5pt]
Email: \email{s2520610@u.tsukuba.ac.jp}, \email{rtakeda@sanken.osaka-u.ac.jp}, \email{aihara@cs.tsukuba.ac.jp}
}

\maketitle

\abstract
Synchronization can be observed in various systems in physics and biology.
The choruses of male frogs are known as an example of biological synchronization in which the well-organized temporal structure, i.e., anti-phase synchronization between neighbors, is realized. 
Given that male frogs produce sounds to advertise their territories to competitors, the dynamics of phases and spatial coordinates should be mutually coupled in the frog choruses. 
In this study, we examined the spatio-temporal dynamics in the choruses consisting of male Japanese tree frogs and other acoustic animals. 
First, we carried out playback experiments using actual frogs and observed that male Japanese tree frog synchronized in anti-phase with the stimuli of a similar frequency but did not synchronize with the stimuli of a much different frequency. 
Second, we modeled the choruses with two species as a system of coupled mobile oscillators and numerically evaluated how the spatio-temporal structure depends on the similarity of call frequencies.
Numerical simulations of the model showed that (1) the two-cluster antisynchronization is established in the same species when the distributions of call frequencies are much different between two species 
and (2) the two-cluster antisynchronization is disturbed when the distributions of call frequencies are similar. 
These results highlight the occurrence of various spatio-temporal patterns  in the proposed model, indicating the importance of repulsive effects with different weights on the variation of the spatio-temporal patterns.
\endabstract

\section{INTRODUCTION}
Synchronization can be observed in various systems in biology and physics.
The examples in biological systems include the flashing of fireflies \cite{buck1966biology} and the circadian rhythms of mammals \cite{yamaguchi2003synchronization}; those in physical systems include Huygens’ clocks \cite{bennett2002huygens} and superconducting Josephson junctions \cite{jain1984mutual}.
The well-organized temporal structures occur in these systems through the interactions among the oscillators. 
To theoretically examine the mechanisms of the synchronization phenomena, the mathematical frameworks such as a phase oscillator model have been widely studied \cite{kuramoto2003}. 
These frameworks have been recently extended to the systems in which the dynamics of the phase and spatial coordinates are bidirectionally coupled \cite{tanaka2007general, aihara2014_SciRepo, okeeffe2017swarmalators}, 
demonstrating rich types of spatio-temporal structures. 

The choruses of male frogs are known as a biological example showing several types of synchronization.  
During the breeding season, male frogs aggregate at aquatic habitats and produce sounds to attract conspecific females \cite{gerhardt2002acoustic}. 
Acoustic interaction between male frogs induces well-organized temporal structures in their choruses \cite{gerhardt2002acoustic, greenfield2021rhythm}. 
For example, laboratory experiments have demonstrated that male Japanese tree frogs (\textit{Hyla japonica}; \textit{Dryophytes japonicus}; \textit{Dryophytes leopardus}) exhibit several types of alternating chorus patterns such as anti-phase synchronization of two individuals \cite{aihara2009PRE}, and clustered anti synchronization and tri-phase synchronization of three individuals \cite{aihara2011complex}; 
field recordings have shown that neighbors of male Japanese tree frogs synchronize in anti-phase with each other \cite{aihara2014_SciRepo, aihara2026arXiv}
, forming two-cluster anti-phase synchronization as a whole. 
Given that the calling behavior of male frogs also plays the role to advertise the territories to conspecific males, the mechanism inherent in the spatio-temporal patterns of frog choruses has been mathematically studied in the context of coupled mobile oscillators \cite{aihara2014_SciRepo}.

In this study, we theoretically examine the spatio-temporal structures in the choruses with multiple species of acoustic animals. 
In such choruses, the acoustic animals may interact not only with conspecific males, but also with the other species via acoustic signals. 
Field recordings have demonstrated that male \textit{H. ebraccata} adjust the timing of their calls in response to calls of the sympatric \textit{H. microcephala}, producing their own calls shortly afterward \cite{schwartz1984interspecific}.
Similarly, laboratory experiments have shown that male \textit{H. microcephala} produce calls shortly after the acoustic stimuli of \textit{H. ebraccata} \cite{schwartz1985Hmicrocephala}.
Combined with the technical difficulty in the sound-source localization and separation in natural environment \cite{aihara2026arXiv}, the spatio-temporal structure and its mechanism in the choruses with multiple species of frogs need further examination in the context of coupled mobile oscillators.

This paper is organized as follows.
First, we recorded the responses of Japanese tree frogs to the acoustic signals of other animals and quantified the synchronization state by calculating the phase difference (Section 2).
Second, we proposed a mathematical model of frog choruses with two species based on the empirical data and numerically evaluated how the spatio-temporal structure changes depending on the similarity of acoustic signals (Section 3).

\begin{figure}[htbp]
    \centering
    \includegraphics[width=1\linewidth]{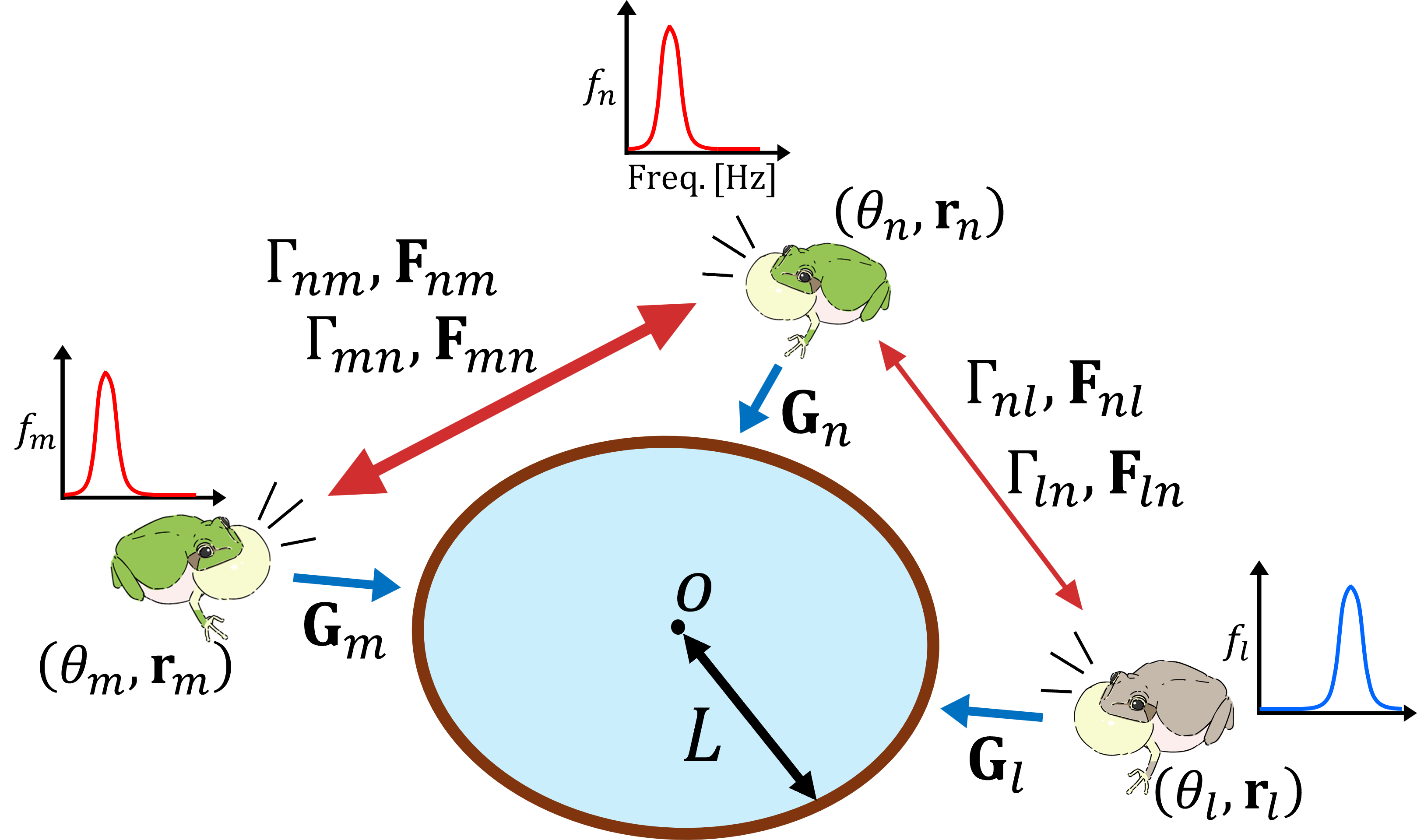}
    \caption{Schematic diagram of our mathematical model for frog choruses with two species. 
    The $n$th frog has a call frequency $f_n$, which is specific to its species.
    The calling time and position of the $n$th frog are modeled by using the calling phase $\theta_n$ and the spatial position $\vb{r}_n$, respectively. 
    The $n$th frog and $m$th frog mutually interact through the functions $\Gamma_{mn}$, $\Gamma_{mn}$, $\vb{F}_{mn}$, and $\vb{F}_{mn}$, which depend on the difference in their call frequencies $f_n$ and $f_m$.
    The function $\vb{G}_n$ represents the effect that attract male frogs towards the edge of a paddy field.
    For simplicity, the geometric shape of the field is assumed to be a circle with radius $L$ centered at the origin $\vb{0}$. }
    \label{fig:system_illust}
\end{figure}

\section{MATHEMATICAL MODELING}

\subsection{Laboratory experiments on interspecific acoustic interactions}

Here, we introduce laboratory experiments supporting the validity of our mathematical modeling. 
Previous studies have revealed that a pair of male Japanese tree frogs tend to synchronize in anti-phase \cite{aihara2009PRE}, demonstrating the importance of intraspecific interaction for the occurrence of synchronized behavior. 
In contrast, this study focuses on interspecific interaction between Japanese tree frogs and the different species of acoustic animals. 
In this study, we have selected two species of acoustic animals, Indian rice frogs (\textit{Fejervarya kawamurai}; \textit{Rana limnocharis}) \cite{djong2011numa} and the crickets (\textit{Modicogryllus siamensis}; \textit{Lepidogryllus siamensis}), and have performed playback experiments to examine the response of actual male Japanese tree frogs towards the calls of these animals.
The reason of this selection is that (1) the inter-call interval of these animals is similar to that of Japanese tree frogs \cite{japanese2003Frog, kim2013tambo}, 
(2) the call frequency of \textit{F. kawamurai} is relatively similar to that of Japanese tree frogs while the frequency of \textit{M. siamensis} is much different \cite{japanese2003Frog, kim2013tambo},
and (3) \textit{F. kawamurai} and \textit{M. siamensis} can be observed in the same fields with Japanese tree frogs.

\begin{figure}[htbp]
    \centering

    \begin{subfigure}[b]{\linewidth}
        \caption{}
        \centering
        \includegraphics[width=0.45\linewidth]{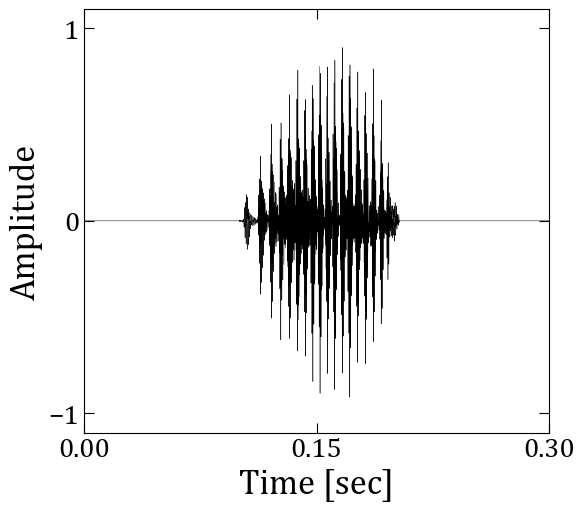}
        \hfill
        \includegraphics[width=0.45\linewidth]{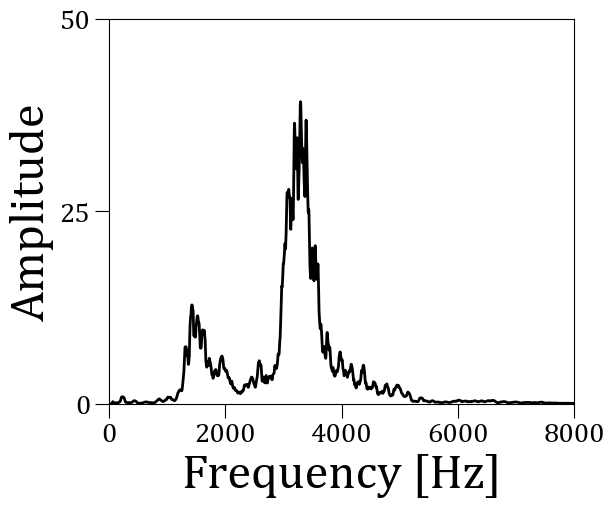}
    \end{subfigure}

    \medskip

    \begin{subfigure}[b]{\linewidth}
        \caption{}
        \centering
        \includegraphics[width=0.45\linewidth]{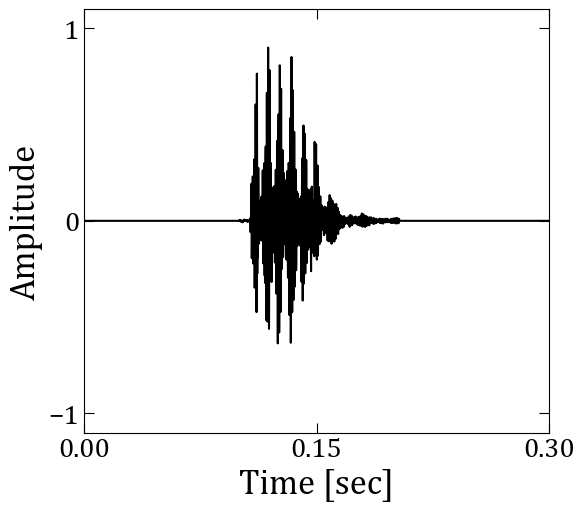}
        \hfill
        \includegraphics[width=0.45\linewidth]{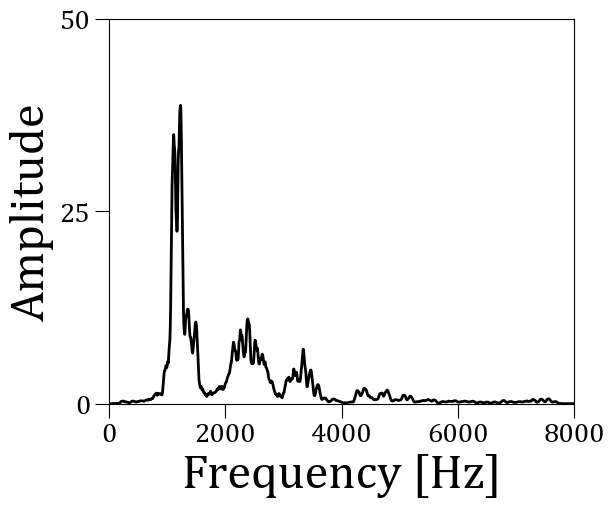}
    \end{subfigure}

    \medskip

    \begin{subfigure}[b]{\linewidth}
        \caption{}
        \centering
        \includegraphics[width=0.45\linewidth]{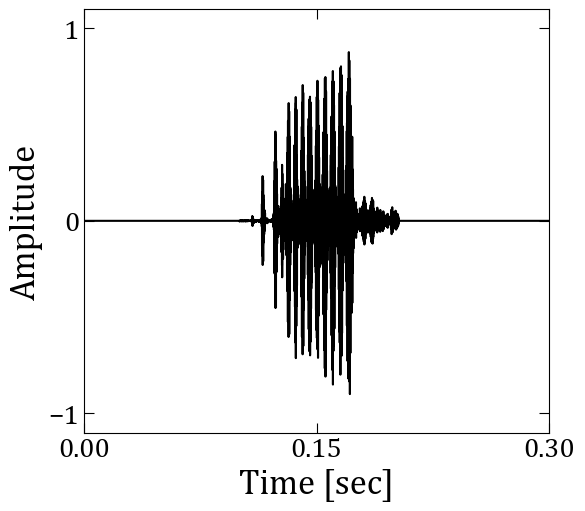}
        \hfill
        \includegraphics[width=0.45\linewidth]{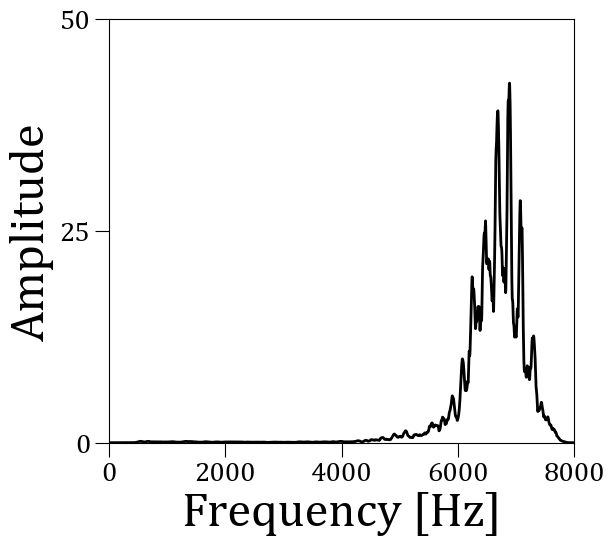}
    \end{subfigure}

    \caption{
        Empirical data on the calls of each species:
        (a) \textit{H. japonica},
        (b) \textit{F. kawamurai}, and
        (c) \textit{M. siamensis}.
        The left panels show the waveforms, whereas the right panels
        show the distributions of call frequencies.
    }
    \label{fig_RealFrog:freqDist_EmpiricalData}
\end{figure}

Prior to the playback experiment, we captured males of \textit{F. kawamurai} and \textit{M. siamensis} at paddy fields in University of Tsukuba and recorded their spontaneous calling behavior in our laboratory by an omnidirectional microphone (JTS, CX-500F) and an audio recorder (Roland, R-44) on 17th, June, 2025. 
Figure \ref{fig_RealFrog:freqDist_EmpiricalData} shows the wave forms and spectrograms of spontaneous calls produced by \textit{F. kawamurai} and \textit{M. siamensis}. 
It was confirmed that (1) \textit{F. kawamurai} periodically produced sounds with the call frequency of 1,200 Hz \cite{djong2011numa} and (2) \textit{M. siamensis} periodically produced sounds with the call frequency of 6,700 Hz. 
To assess the similarity of acoustic signals, we calculated the difference between the dominant peaks of the spectrograms for each pair of animal species and treated it as the difference of call frequency \cite{allen2022accositicdifference}. 
The difference of the frequency between Japanese tree frog and \textit{F. kawamurai} was approximately 2,100 Hz while that between Japanese tree frog and \textit{M. siamensis} was approximately 3,400 Hz.
Thus, the call frequency of Japanese tree frog was relatively similar to that of \textit{F. kawamurai}, compared to the frequency of \textit{M. siamensis}. 

Next, we performed the playback experiments to examine the response of male Japanese tree frogs towards the calls of other acoustic animals.
First, we captured male Japanese tree frogs at paddy fields in University of Tsukuba and put each male in a small meshed cage ($140\ \mathrm{(width)}\times100\ \mathrm{(depth)}\times145\ \mathrm{(height)}$ mm).
Second, the periodic calls of \textit{F. kawamurai} and \textit{M. siamensis} were broadcast from a loudspeaker (audio-technica, AT-MSP56TV) with the same maximum amplitude when the calling behavior of an actual frog was detected. 
Third, we recorded spontaneous of male Japanese tree frogs by an omnidirectional microphone (JTS, CX-500F) and an audio recorder (Roland, R-44). 
These experiments were performed on 2nd, July, 2025 and 3rd, July, 2025.

\begin{figure}[htbp]
    \centering

    \begin{subfigure}[t]{0.45\linewidth}
        \vspace{0pt}
        \renewcommand{\thesubfigure}{a1}
        \caption{}
        \centering
        \includegraphics[height=3.5cm, keepaspectratio]{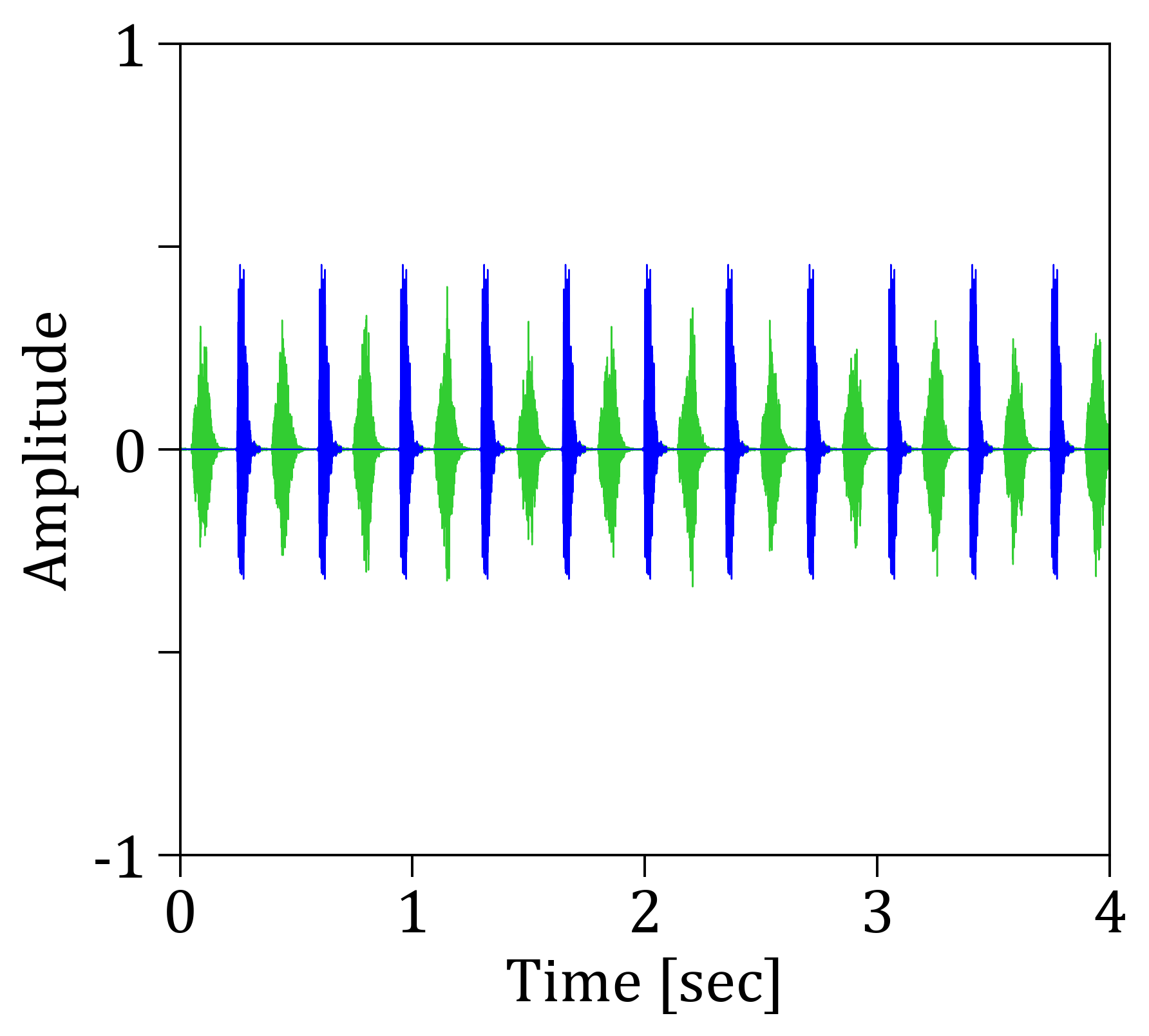}
        \label{fig_RealFrog:numa2ama}
    \end{subfigure}
    \hspace{6mm}
    \begin{subfigure}[t]{0.4\linewidth}
        \vspace{0pt}
        \renewcommand{\thesubfigure}{a2}
        \caption{}
        \centering
        \includegraphics[height=3.5cm, keepaspectratio]{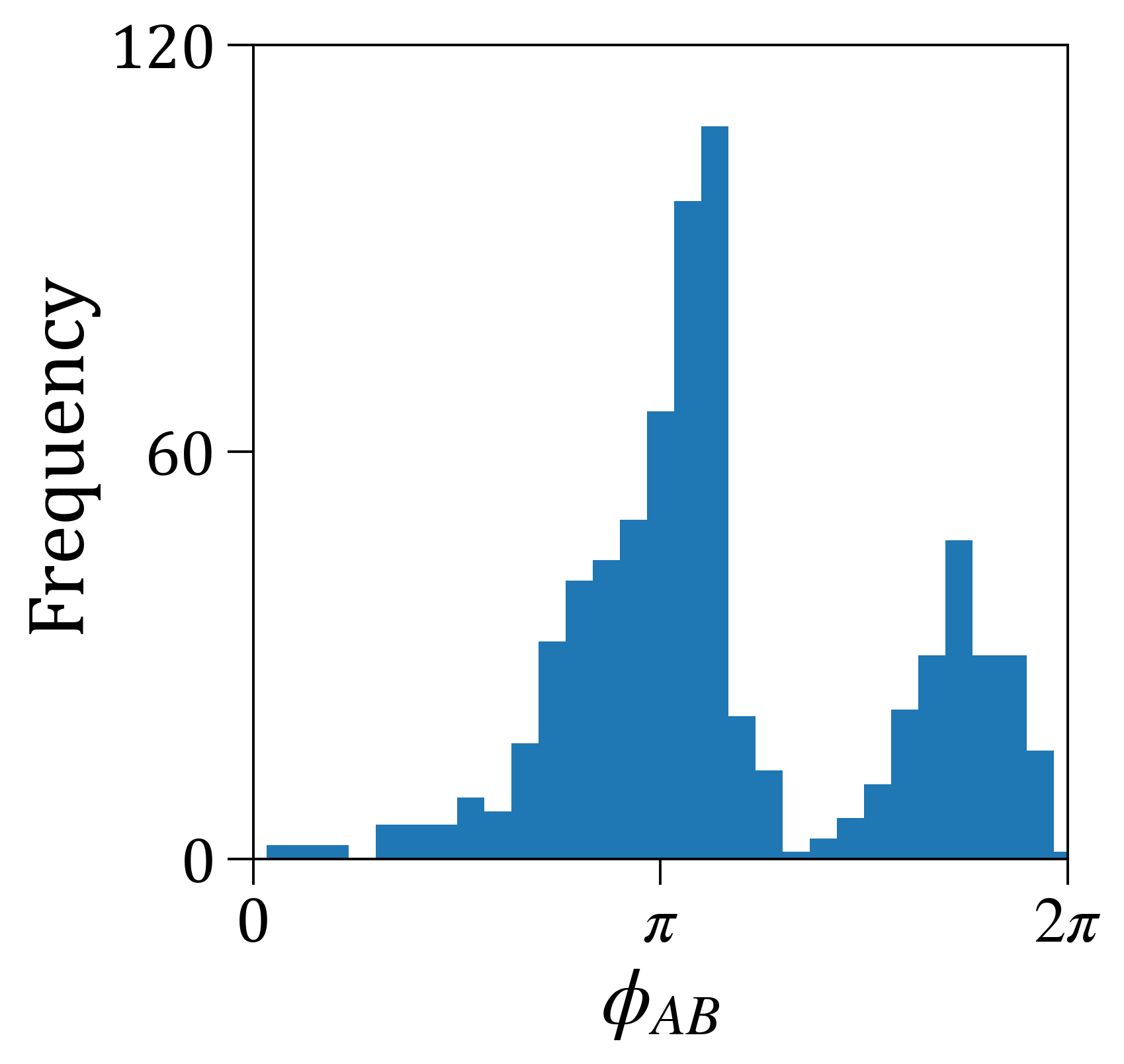}
        \label{fig_RealFrog:numa2ama_histgram}
    \end{subfigure}
    
    \medskip
    
    \begin{subfigure}[t]{0.45\linewidth}
        \vspace{0pt}
        \renewcommand{\thesubfigure}{b1}
        \caption{}
        \centering
        \includegraphics[height=3.5cm, keepaspectratio]{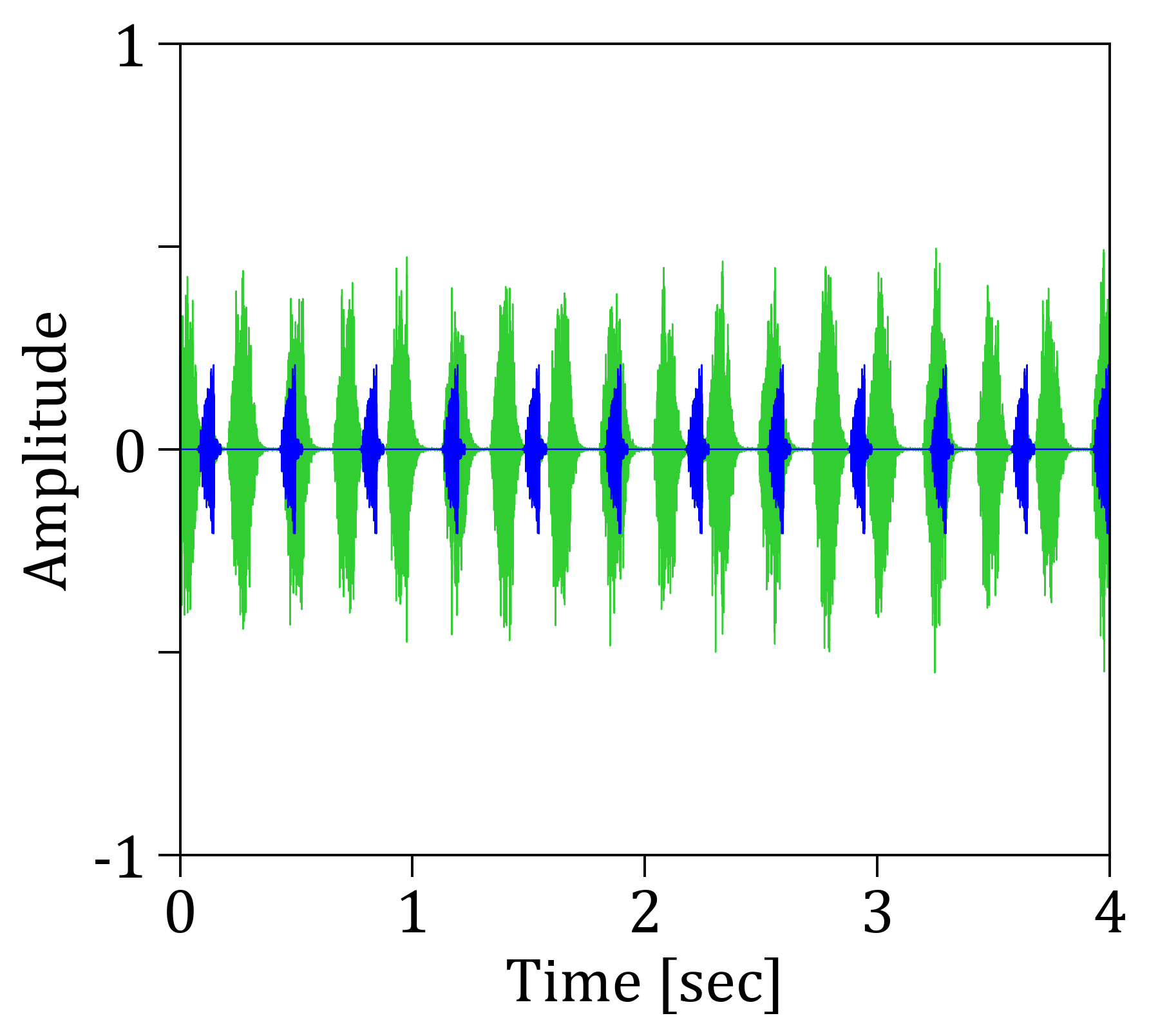}
        \label{fig_RealFrog:tanbo2ama}
    \end{subfigure}
    \hspace{6mm}
    \begin{subfigure}[t]{0.4\linewidth}
        \vspace{0pt}
        \renewcommand{\thesubfigure}{b2}
        \caption{}
        \centering
        \includegraphics[height=3.5cm, keepaspectratio]{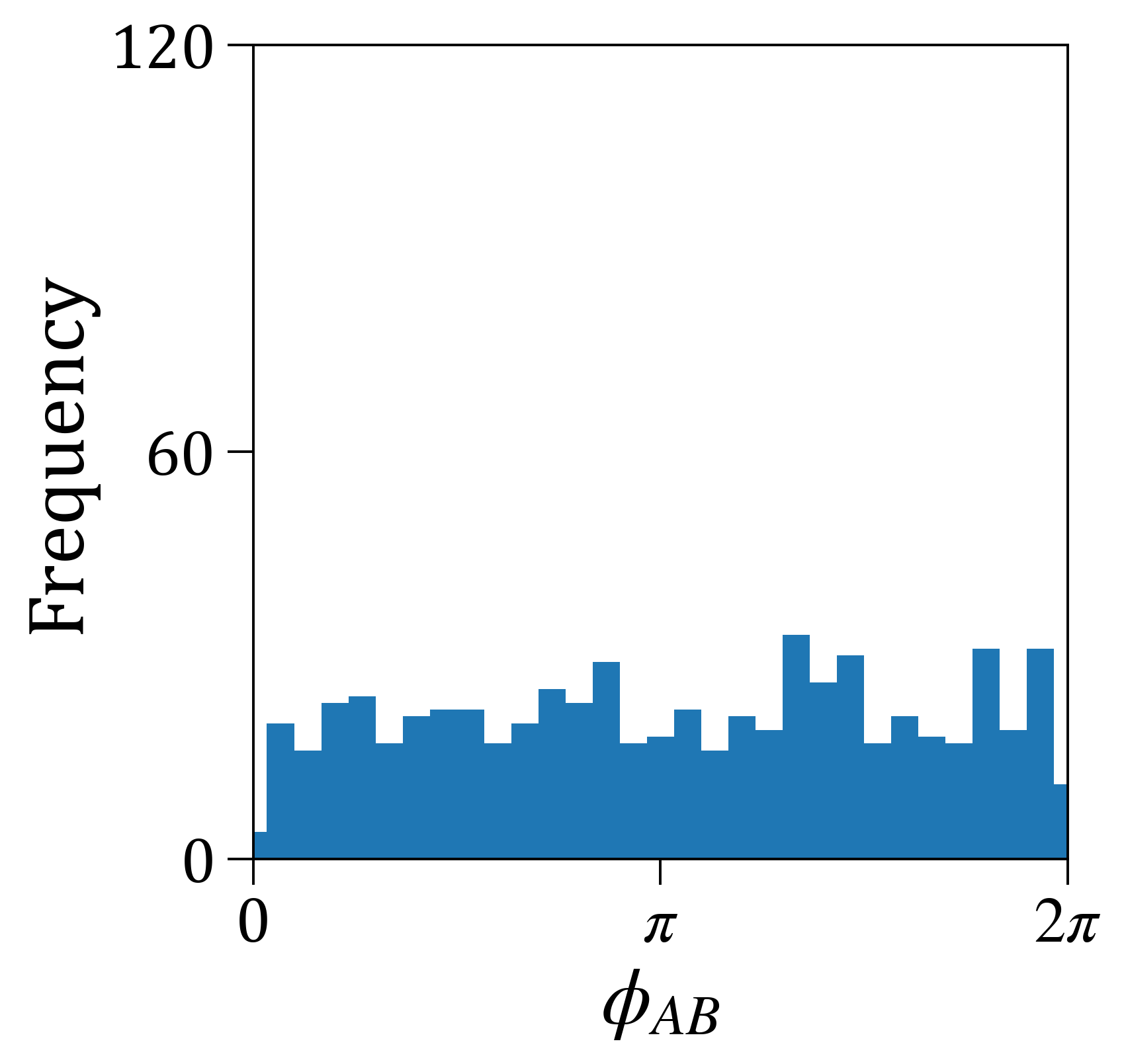}
        \label{fig_RealFrog:tambo2ama_histgram}
    \end{subfigure}
    
\caption{ 
  Empirical data on the responses of Japanese tree frogs towards acoustic signals of other animals: (a) \textit{F. kawamurai} and (b) \textit{M. siamensis}.
  In left panels, the green and blue lines represent the waveforms of the Japanese tree frog and sound stimuli, respectively.
  A Japanese tree frog called alternately with the stimuli of \textit{F. kawamurai} while he did not synchronize with the stimuli of \textit{M. siamensis}. 
  Right panels show the histogram of the phase difference $\phi_{AB}$ between a Japanese tree frog and other animals. 
  There is an obvious peak around $\pi$ for the stimuli of \textit{F. kawamurai} while there is no peak for the stimuli of \textit{M. siamensis}.}
   \label{figs_RealFrog:calling behaivior}
\end{figure}

Figure \ref{figs_RealFrog:calling behaivior}(\subref{fig_RealFrog:numa2ama}) and \ref{figs_RealFrog:calling behaivior}(\subref{fig_RealFrog:tanbo2ama}) show the response of male Japanese tree frogs to each sound stimulus.
Here, we separated audio signals of actual frogs based on the custom-made loop-back system for signal processing by using unprocessed audio recordings and playback sounds (see Appendix A for details). 
It was demonstrated that Japanese tree frogs called alternately in anti-phase with the stimuli of \textit{F. kawamurai} while they did not show any synchronized behavior towards the stimuli of \textit{M. siamensis}.
Next, we quantified these responses of actual frogs in the context of synchronization phenomena by using the phase difference as follows \cite{aihara2011complex}:
\begin{align}
    \phi_{AB} = 2\pi \frac{t_B^{l} - t_A^{k}}{t_A^{k+1} - t_A^{k}} \label{eq: phase differece}.
\end{align}
Here $ t_A^{k}$ represents the time of the $k$th call produced by the actual frog; $t_B^{l}$ represents the time of the $l$th stimulus broadcast from the loudspeaker. 
Given the periodicity in the calling behavior of Japanese tree frogs \cite{aihara2009PRE} as well as the sound stimuli, the phase difference was calculated only when $t_A^{k} \le t_B^{l} \le t_A^{k+1}$ and $t_A^{k+1} - t_A^{k} < 0.45$ sec.
When the frog exhibits nearly anti-phase synchronization towards the stimuli, $\phi_{AB}$ is expected to be close to $\pi$.
Figures \ref{figs_RealFrog:calling behaivior}(\subref{fig_RealFrog:numa2ama_histgram}) and \ref{figs_RealFrog:calling behaivior}(\subref{fig_RealFrog:tambo2ama_histgram}) show the histograms of the phase difference $\phi_{AB}$. 
As for the response towards the stimuli of \textit{F. kawamurai}, the distribution of the phase difference was significantly localized around $\pi$ (Fig. \ref{figs_RealFrog:calling behaivior}(\subref{fig_RealFrog:numa2ama_histgram}); $p<0.001$ with Rayleigh test). 
In contrast, as for the stimuli of \textit{M. siamensis}, the distribution of the phase difference was not significantly localized around a specific value (Fig. \ref{figs_RealFrog:calling behaivior}(\subref{fig_RealFrog:tambo2ama_histgram}); $p \approx 0.58$ with Rayleigh test).
Thus, our empirical data have indicated that the similarity in frequency of acoustic signals between two species influences the strength of interspecific interactions, thereby inducing the synchronized behavior.


\subsection{Mathematical modeling of interspecific communication in male frogs}

We propose a mathematical model describing the interspecific and intraspecific communication in the choruses of male frogs.
Whereas a single male frog calls periodically, a pair of the male frogs interact through sounds within the same species \cite{aihara2014_SciRepo, aihara2009PRE}. 
Furthermore, male frogs produce the calls with different frequencies depending on species \cite{japanese2003Frog} (see the right panels of Fig. \ref{fig_RealFrog:freqDist_EmpiricalData} for example). 
Combined with the empirical result of Section 2.1, we hypothesize that the difference in the call frequencies affects the strength of the interaction.
We model such an acoustic communication between different species of animals by extending the mathematical model of coupled mobile oscillators \cite{aihara2014_SciRepo} as follows (Fig. \ref{fig:system_illust}):
\begin{align}
    \dot{\theta}_n &= \omega_n + \sum_{m\ne n} \Gamma_{mn}(\theta_m-\theta_n,\ \vb{r}_m-\vb{r}_n,\ f_m,\ f_n), \label{phase dynamics} \\
    \dot{\vb{r}}_n &= \sum_{m\ne n} \vb{F}_{mn}(\theta_m- \theta_n,\ \vb{r}_m-\vb{r}_n,\ f_m,\ f_n) + \vb{G}_{n}(\vb{r}_n) \label{spatio dynamics}.
\end{align}
Here $\theta_n\in\mathbb{S}^1\ (n=1,\dots, N)$ represents the phase of the calls produced by the $n$th frog  {\cite{aihara2014_SciRepo, aihara2009PRE}}; the vector $\vb{r}_n\in\mathbb{R}^2$ represents the position of the $n$th frog {\cite{aihara2014_SciRepo}}. 
The parameter $\omega_n$ is the intrinsic angular velocity of the $n$th frog; $f_n$ represents the distribution of the call frequency in the $n$th frog.
The functions $\Gamma_{mn}(\theta_m-\theta_n,\ \vb{r}_m-\vb{r}_n,\ f_m,\ f_n) \in \mathbb{R} $ and $\vb{F}_{mn}(\theta_m- \theta_n,\ \vb{r}_m-\vb{r}_n,\ f_m,\ f_n)\in\mathbb{R}^2$ $(n,\ m = 1, 2, \dots, N\ \mathrm{and}\ n \ne m)$ describe the effects from the $m$th frog to the $n$th frog depending on their acoustic communication {\cite{aihara2014_SciRepo}}. 
It should be noted that these functions are assumed to depend on the distributions of their call frequencies, $f_m$ and $f_n$, which is the extension from the previous study {\cite{aihara2014_SciRepo}}. 
The function $\vb{G}_{n}(\vb{r}_n)\in\mathbb{R}^2$ represents the effect by which male frogs aggregate along the edge of a paddy field for breeding {\cite{aihara2014_SciRepo}}.

Next, we determine the specific forms for three functions $\Gamma_{mn}(\theta_m-\theta_n,\ \vb{r}_m-\vb{r}_n,\ f_m,\ f_n),\ \vb{F}_{mn}(\theta_m- \theta_n,\ \vb{r}_m-\vb{r}_n,\ f_m,\ f_n), \mathrm{and}\ \vb{G}_{n}(\vb{r}_n)$ in Equations (\ref{phase dynamics}) and (\ref{spatio dynamics}) on the basis of the previous study as follows {\cite{aihara2014_SciRepo, aihara2009PRE}}:
\begin{align}
    \Gamma_{mn}(\theta_m-\theta_n,\ \vb{r}_m-\vb{r}_n,\ f_m,\ f_n) \notag\\
    =  -K_{mn}(\vb{r}_m-\vb{r}_n,\ f_m,\ f_n) \sin(\theta_m-\theta_n), \label{Gamma}\\
        \vb{F}_{mn}(\theta_m- \theta_n,\ \vb{r}_m-\vb{r}_n,\ f_m,\ f_n) \notag \\
    =  -K_{mn}(\vb{r}_m-\vb{r}_n,\ f_m,\ f_n) \qty[1-\cos(\theta_m - \theta_n)] \vb{e}_{mn}, \label{F}\\
    \vb{G}(\vb{r}_n) = \alpha(L - \abs{\vb{r}_n})^3\vb{e}_n. \label{G(r)}
\end{align}
Equations (\ref{Gamma}) and (\ref{F}) provide the same framework with our previous study {\cite{aihara2014_SciRepo, aihara2009PRE}}, except that the coupling strength depends on the distributions of call frequencies ($f_m$ and $f_n$). 
In Equation (\ref{F}), $\vb{e}_{mn}$ is the unit vector from the $n$th frog to the $m$th frog (i.e., $\vb{e}_{mn} = (\vb{r}_m - \vb{r}_n)/ \abs{\vb{r}_m - \vb{r}_n)}$).
Eventually, these terms are expected to reproduce anti-phase synchronization between neighbors and then achieve the maintenance of frogs' territories in the aggregation {\cite{aihara2014_SciRepo}}. 
Equation (\ref{G(r)}) describes the attraction of male frogs towards the edge of a breeding site (a paddy filed) where they attempt to attract conspecific females by calling, which is consistent with the framework of the previous study {\cite{aihara2014_SciRepo}}.
Here, the parameter $L$ represents the radius of a circular breeding site; 
$\alpha$ represents the strength of the attractive effect from the $n$th frog to the edge of the breeding site.
$\vb{e}_{n}$ is a unit vector from the center of the breeding site $\vb{0}$ to the position of the $n$th frog $\vb{r}_n$ (i.e., $\vb{e}_{n} = \vb{r}_n / \abs{\vb{r}_n}$). 
The difference is that a cubic term is used instead of a linear term {\cite{aihara2014_SciRepo}} in this Equation to avoid the rapid movement near the edge that is unrealistic in the behavior of actual frogs.

\begin{figure}[htbp]
    \centering
    \includegraphics[width=0.6\linewidth]{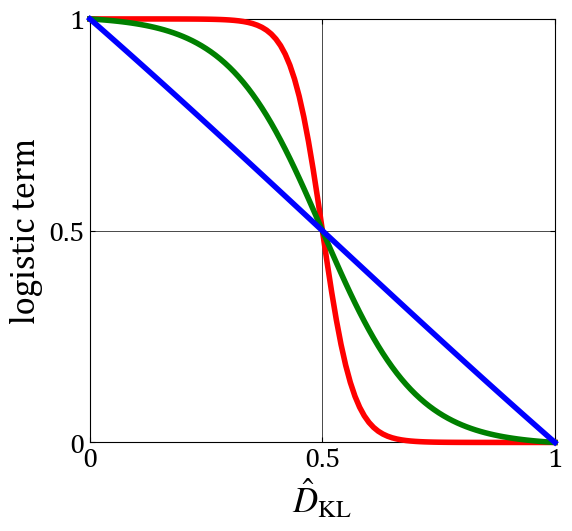}
    \caption{Variation of the logistic term reproduced by Equation (\ref{K_mn}). 
    The solid red, green, and blue lines represent the shapes of the logistic term on the assumption of $c_{\mathrm{KL}}=1$, $10$, and $30$, respectively. 
    The shape becomes nearly linear as $c_{\mathrm{KL}}$ approaches zero.
    The parameters $a$ and $b$ allow us to restrict the range of the logistic term between 0 and 1.}
    \label{fig:c_KL}
\end{figure}

The function, $K_{mn}(\vb{r}_m-\vb{r}_n,\ f_m,\ f_n)$ in Equations (\ref{Gamma}) and (\ref{F}), represents the coupling strength from the $m$th frog to the $n$th frog.
Here we assume that $K_{mn}$ is determined for each pair of male frogs by two factors: (1) the distance between the frogs and (2) the similarity of call frequencies. 
The first assumption is consistent with the previous study \cite{aihara2014_SciRepo}: namely, $K_{mn}(\vb{r}_m-\vb{r}_n,\ f_m,\ f_n)$ is inversely proportional to the square of the distance following the inverse-square law for sound propagation.
The second assumption is based on our empirical results of Figures \ref{fig_RealFrog:numa2ama_histgram} and \ref{fig_RealFrog:tambo2ama_histgram} indicating that the difference of call frequencies can modulate the occurrence of synchronized behavior.
Consequently, we formulate the coupling strength $K_{mn}(\vb{r}_m-\vb{r}_n,\ f_m,\ f_n)$ as follows: 
\begin{align}
    & K_{mn}(\vb{r}_m-\vb{r}_n,\ f_m,\ f_n) \notag\\
    &= \frac{1}{\abs{\vb{r}_m-\vb{r}_n}^2} \qty{ \frac{a}{ 1 + \exp \qty[  c_{\mathrm{KL}} \qty( \hat{D}_{\mathrm{KL}}(f_m,f_n) - \frac{1}{2} ) ] } - b }, \label{K_mn}
\end{align}
where,
\begin{align}
    {\hat{D}_{\mathrm{KL}}(f_m, f_n) = D_{\mathrm{KL}}(f_m, f_n) / D_{\mathrm{KL, max}}~.} \label{eq:normalized_DKL}
\end{align}
Here, we quantified the similarity in call frequencies by the Kullback–Leibler divergence (KL divergence, $D_{\mathrm{KL}}(f_m,f_n)$).
Given that $D_{\mathrm{KL}}$ does not have an upper limit, we introduce the normalized KL divergence $\hat{D}_{\mathrm{KL}}(f_m, f_n)$. 
In Equation (\ref{eq:normalized_DKL}), $D_{\mathrm{KL, max}}$ is the maximum value of $D_{\mathrm{KL}}$ for the distributions of call frequencies that are possible in actual acoustic animals in our field site. 
Subsequently, $\hat{D}_{\mathrm{KL}}(f_m,\ f_n)$ is equal to the lower bound of 0 when $f_m$ and $f_n$ are identical, inducing the strongest coupling between the focal pair of frogs due to the logistic function in Equation(\ref{K_mn}). 
In contrast, $\hat{D}_{\mathrm{KL}}(f_m,\ f_n)$ is equal to the upper bound of 1 when $f_m$ and $f_n$ are much different, inducing the weakest coupling between the focal pair. 
Note that the parameters $a$, $b$ and $c_{\mathrm{KL}}$ in Equation (\ref{K_mn}) allow us to change the shape of the logistic function (see Fig. \ref{fig:c_KL}).

\subsection{Numerical simulation}

First, we fixed some of the parameters in our model (Equation (\ref{phase dynamics})--(\ref{K_mn})).
Following empirical data in our previous study {\cite{aihara2014_SciRepo}}, the total number of male frogs and the intrinsic angular velocity of each frog are set as $N=20$ and $\omega_n=8\pi\ \mathrm{rad/s}$, respectively.
These frogs are divided equally into two species, with 10 frogs per species.
Then, we set the radius of the paddy field as $L=40$ m. 
This value is larger than that of the previous study \cite{aihara2014_SciRepo}, inducing the longer inter-frog distance around $13$m.
On the other hand, recent studies have succeeded in quantifying the inter-frog distance in natural environment and have indicated that $13$m is close to the upper limit of the distance between neighboring callers.
Therefore, we consider that $L=40$m is relatively larger but still reproduces the spatial coordinates of frogs that are consistent with the field observation.
The parameter $c_{\mathrm{KL}}$ in Equation (\ref{K_mn}) is difficult to be estimated from empirical data.
Hence, we examine two cases: $c_{\mathrm{KL}}=10$ and $30$. 
The parameters $a$ and $b$ act as the coefficients for min-max normalization, derived from the maximum and minimum values of the original logistic function at $\hat{D}_{\mathrm{KL}}=0$ and $\hat{D}_{\mathrm{KL}}=1$, respectively.
Accordingly, the parameters $a$ and $b$ in Equation (\ref{K_mn}) are automatically determined for a given value of $c_{\mathrm{KL}}$.
Due to the extension of our model in Equation (\ref{G(r)}) (i.e., the novel use of the cubic term for modeling the attraction of males to the paddy field), the coefficient $\alpha$ is fixed at a smaller value of $0.01$ than in the previous study \cite{aihara2014_SciRepo}.

\begin{figure}[htbp]
    \centering
    \begin{subfigure}[t]{0.4\linewidth}
        \caption{}
        \centering
        \includegraphics[height=3cm, keepaspectratio]{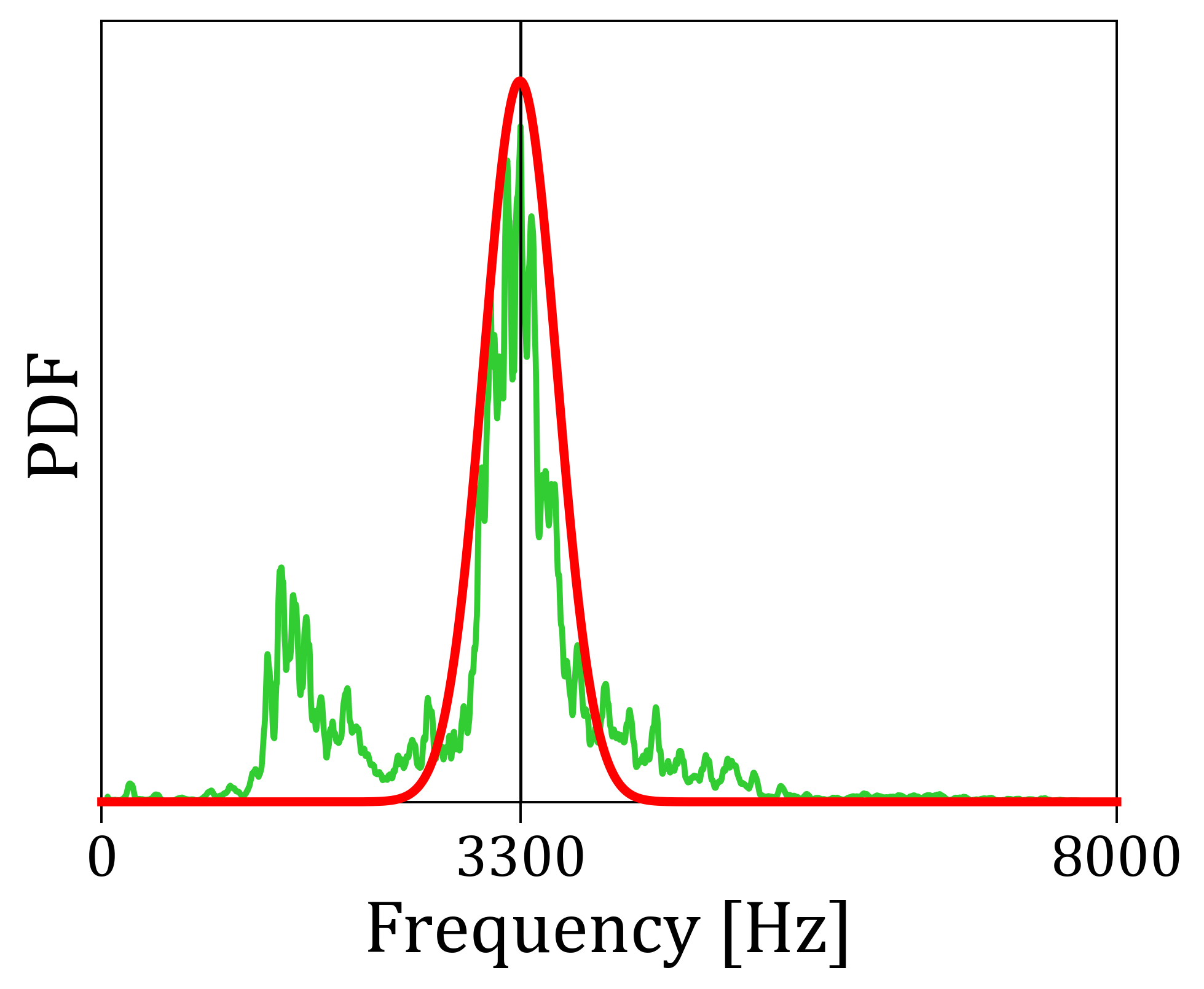}
        \label{fig: modeling call freq. ama}
    \end{subfigure}
    \hspace{1cm}
    \begin{subfigure}[t]{0.4\linewidth}
        \caption{}
        \centering
        \includegraphics[height=3cm, keepaspectratio]{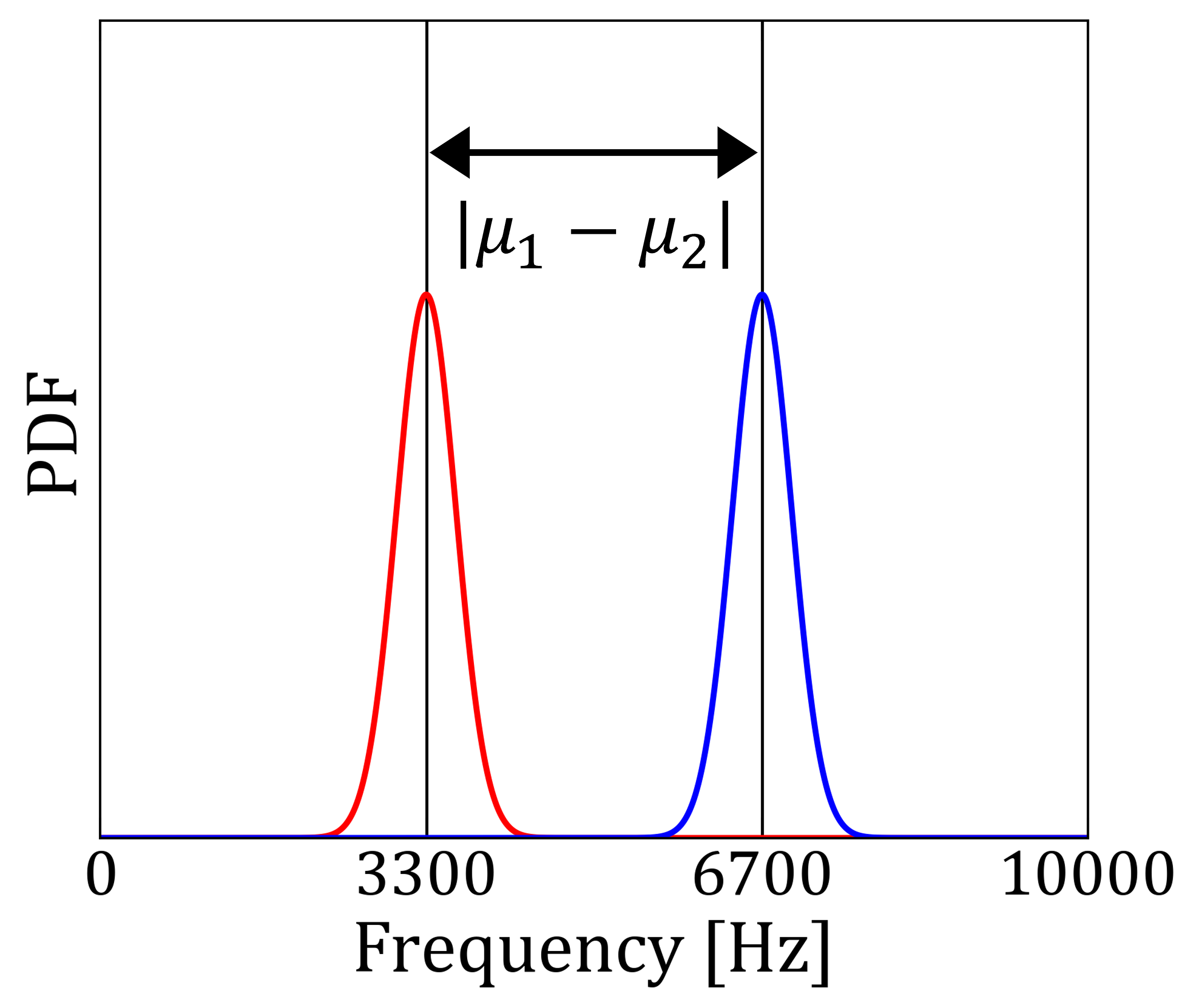}
        \label{fig: normal distributions}
    \end{subfigure}\\
    \begin{subfigure}[t]{0.4\linewidth}
        \caption{}
        \centering
        \includegraphics[height=3.5cm, keepaspectratio]{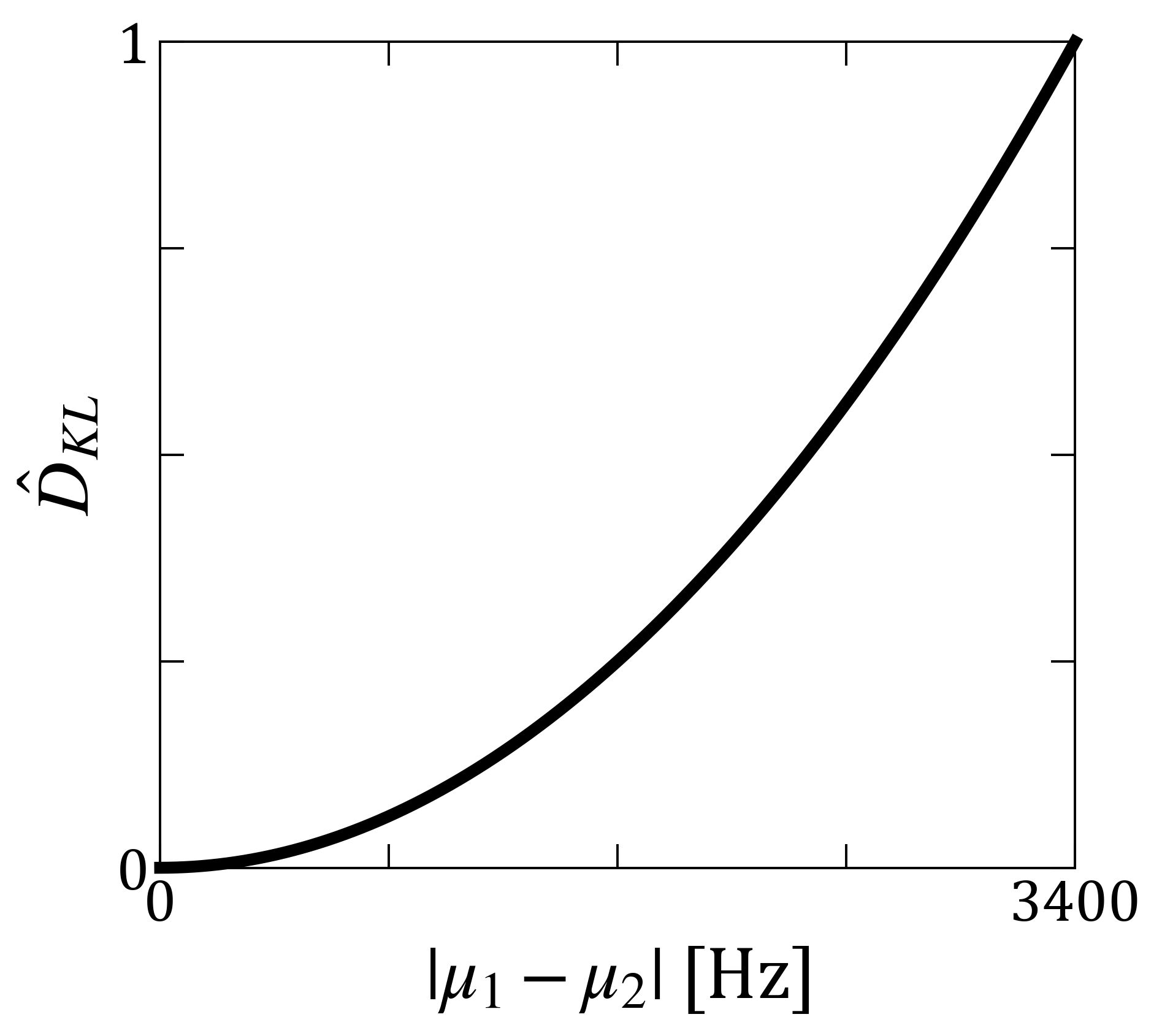}
        \label{fig:KLdiv for normal distributions}
    \end{subfigure}

    \caption{Quantification of the difference in call frequencies using the KL divergence $\hat{D}_{\mathrm{KL}}$.
    (a) The distribution of call frequency for Japanese tree frog. 
    While the green line represents the distribution calculated from empirical data, 
    the red line represents the distribution fitted by a normal distribution with a mean of 3,300 Hz and a standard deviation of 300 Hz. 
    (b) Examples of the models on the distributions of call frequencies for two species. 
    The solid red and blue lines represent the models of call frequency for species 1 and species 2, respectively. 
    The difference between the distributions is quantified by $\abs{\mu_1-\mu_2}$. 
    (c) Dependence of $\hat{D}_{\mathrm{KL}}$ on the difference in call frequencies $\abs{\mu_1-\mu_2}$. 
    $\hat{D}_{\mathrm{KL}}$ is the lower bound of 0 at $\abs{\mu_{1}-\mu_{2}}=0$ Hz and monotonically increases as $\abs{\mu_{1}-\mu_{2}}$ increases, 
    reaching the upper bound of 1 at $\abs{\mu_{1}-\mu_{2}}=3{,}400$ Hz.}
\end{figure}

Next, we set the distribution of call frequency for each species. 
In general, acoustic animals produce sounds at species-specific frequencies \cite{japanese2003Frog, jaiswara2013cricket}. 
Our empirical data demonstrate that the call frequencies of Japanese tree frog and \textit{F. kawamurai} show dominant peaks at approximately 3,300 Hz and 1,200 Hz, respectively (Fig. \ref{fig_RealFrog:freqDist_EmpiricalData}). 
To model this characteristic, we use the normal distribution $N(\mu_s, \sigma_{s}^2)$ with the mean $\mu_s$ ($s=1, 2$) and its standard deviation $\sigma_s$ for each species $s$ (Figs. \ref{fig: modeling call freq. ama} and \ref{fig: normal distributions}).
Furthermore, the difference in the standard deviations around the dominant peaks is small both between Japanese tree frog and \textit{F. kawamurai}, and between Japanese tree frog and \textit{M. siamensis} (see Appendix B for details). Hence, we set $\sigma_s$ to the same value in this study.
Consequently, we describe the difference of call frequencies as a parameter $\abs{\mu_1-\mu_2}$. 

Given that the KL divergence does not have the upper limit, we normalize $D_{\mathrm{KL}}$ on the basis of our empirical data.
Playback experiments have demonstrated that the Japanese tree frog does not synchronize with the acoustic stimuli of \textit{M. siamensis}, and the difference in their call frequencies is approximately $3{,}400~\mathrm{Hz}$ (see Section 2.1). 
Based on this result, we assume that the difference of $\abs{\mu_{1}-\mu_{2}} = 3{,}400~\mathrm{Hz}$ is sufficient to eliminate the interaction.
Accordingly, we have calculated the KL divergence $D_{\mathrm{KL,max}}$ (Eq. (\ref{eq:normalized_DKL}) by using $\abs{\mu_{1}-\mu_{2}} = 3{,}400~\mathrm{Hz}$ and used it for the normalization of $D_{\mathrm{KL}}$.
Figure \ref{fig:KLdiv for normal distributions} shows how the normalized KL divergence $D_{\mathrm{KL}}$ depends on the difference in call frequencies $\abs{\mu_{1}-\mu_{2}}$: $\hat{D}_{\mathrm{KL}}$ is the lower bound of 0 at $\abs{\mu_{1}-\mu_{2}}=0$ Hz and monotonically increases as $\abs{\mu_{1}-\mu_{2}}$ increases, reaching the upper bound of 1 at $\abs{\mu_{1}-\mu_{2}}=3{,}400$ Hz.

Finally, we set the initial condition of numerical simulations. 
Our model describes the dynamics of two factors: the positions and phases of male frogs (Eq. (\ref{phase dynamics}) and \ref{spatio dynamics}).
For the positions, we assume that frogs have already aggregated near the paddy field and then started moving around. 
Based on this assumption, the frogs are assumed to randomly position within the range $L-1 < |\vb{r}_n| < L+10$. 
Then we assume that the initial phases are randomly set within $[0, 2\pi)$.

\subsection{Order parameter}

The previous study on male Japanese tree frogs {\cite{aihara2014_SciRepo}} has revealed that the pairs of neighbors synchronize in anti-phase. 
To detect such an anti-phase synchronization between neighbors within the same species, we introduce the following order parameter: 
\begin{align}
    R_{s} = -\frac{1}{N_{s}}\sum_{i=1}^{N_{s}}\cos(\theta_{i+1}-\theta_{i})  \label{eq:Rs}, 
\end{align}
where $N_s$ represents the number of frogs in each species. 
The index $i$ represents the ID of male frogs in the same species; 
this ID is assigned to each male in ascending order in a counterclockwise direction along the edge of the paddy field. 
It should be noted that the $1$st frog and the $N_s$th frog corresponds to a neighboring pair (i.e.  $N_s + 1 \equiv 1$) because the circular paddy field is assumed in this study. 
Consequently, the anti-phase synchronization in a specific pair of neighbors gives a positive contribution to the order parameter $R_{s}$ as $-\cos(\pi)=1$; 
$R_s$ takes the maximum value of 1 when all the pairs of neighboring frogs synchronize in anti-phase.

\section{RESULT}
\subsection{Spatio-temporal structure}

\begin{figure}[htbp]
    \centering
    \begin{subfigure}[t]{0.48\linewidth}
        \vspace{0pt}
        \caption{}
        \centering
        \includegraphics[height=3.5cm, keepaspectratio]{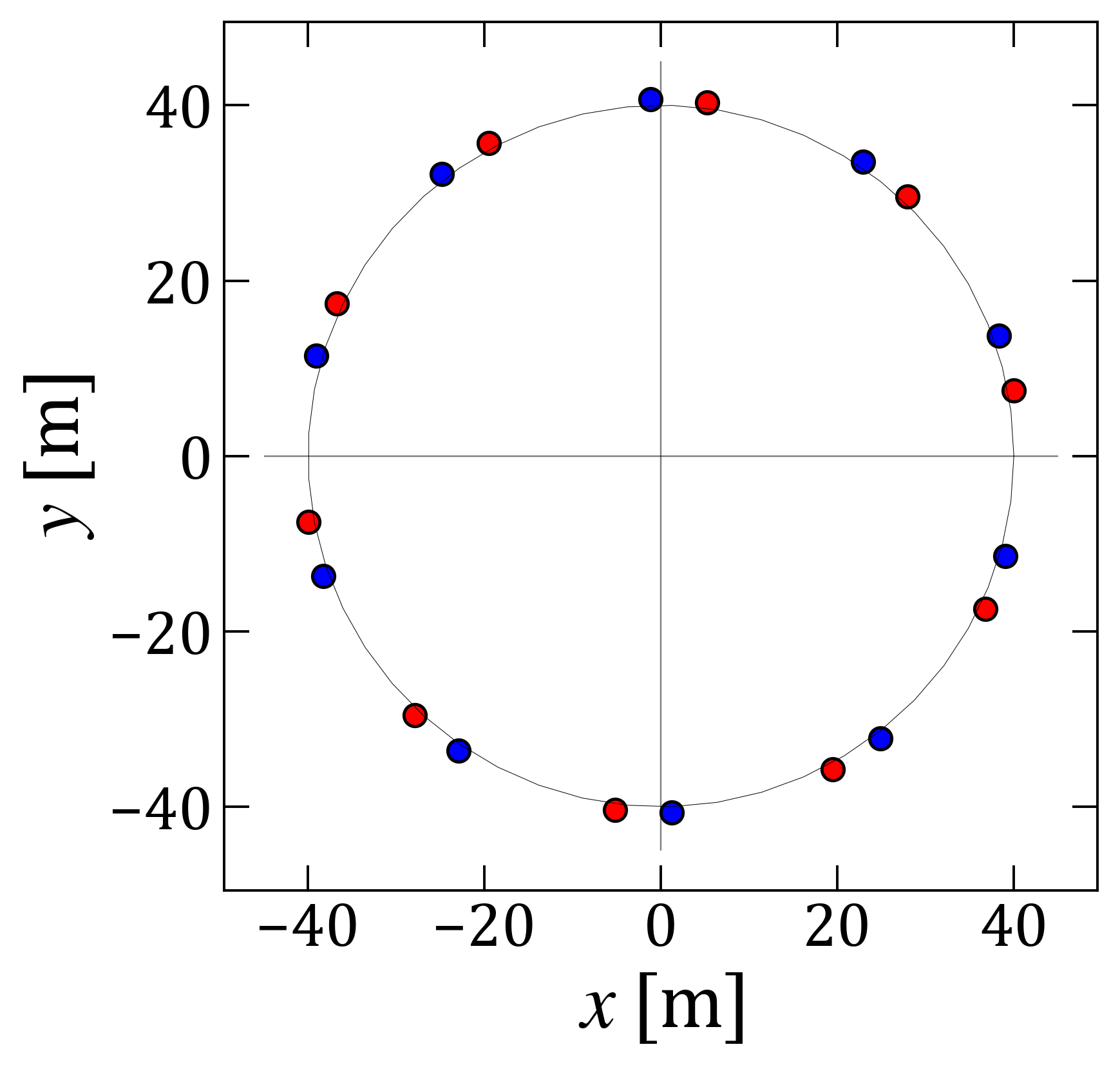}
        \label{fig:KL1_spatioDist}
    \end{subfigure}\\

    \begin{subfigure}[t]{\linewidth}
        \vspace{0pt}
        \caption{}
        \centering
        \includegraphics[height=3.5cm, keepaspectratio]{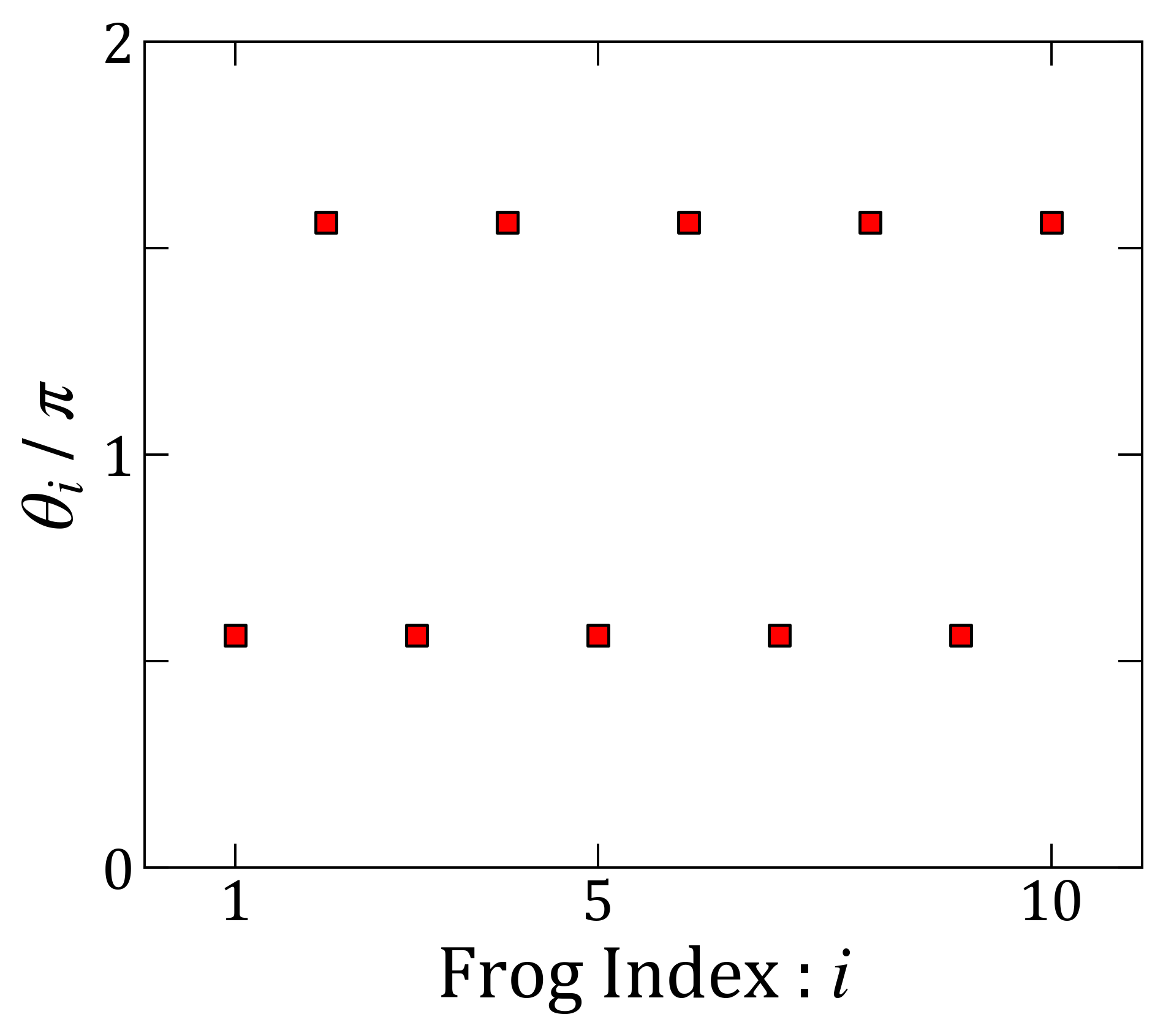}
        \hfill
        \includegraphics[height=3.5cm, keepaspectratio]{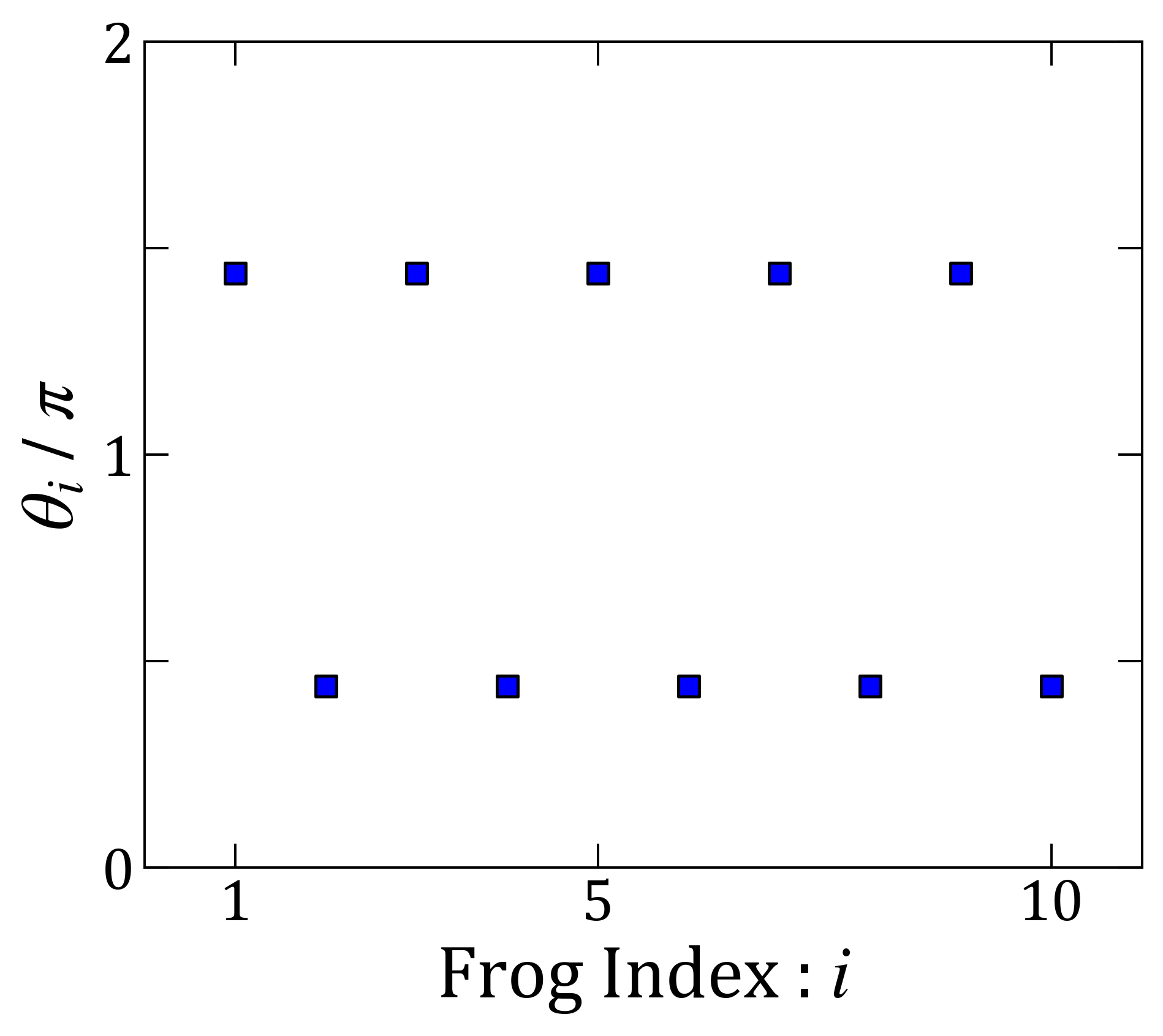}   
        \label{fig:KL1_phaseDist}        
    \end{subfigure}
    
    \caption{Numerical simulation of our model on the spatio-temporal structure at $\hat{D}_{\mathrm{KL}}=1$.  
    (a) The position of male frogs around the breeding site. 
    The males of each species are positioned along the edge of the paddy field at the same inter-frog distance.
    (b) Two-cluster antisynchronization in a frog chorus within the same species. 
    The pairs of neighboring frogs show anti-phase synchronization, forming the two cluster state.}
    \label{figs:KL1_steadyDist}
\end{figure}

\begin{figure}[htbp]
    \centering
    \begin{subfigure}[t]{0.48\linewidth}
        \vspace{0pt}
        \caption{}
        \centering
        \includegraphics[height=3.5cm, keepaspectratio]{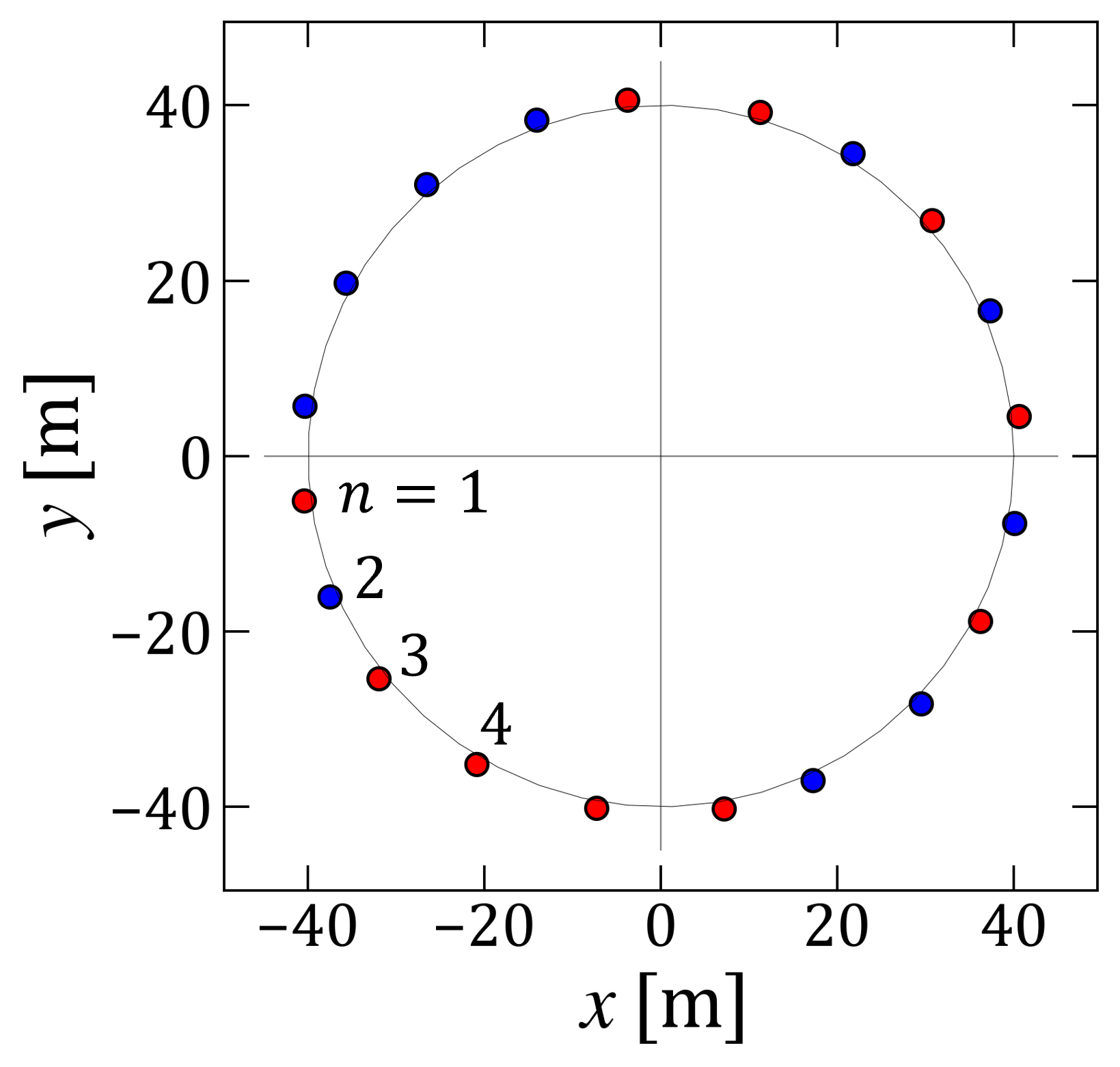}
        \label{fig:KL05_spatioDist}
    \end{subfigure}
    \begin{subfigure}[t]{0.48\linewidth}
        \vspace{0pt}
        \caption{}
        \centering
        \includegraphics[height=3.5cm, keepaspectratio]{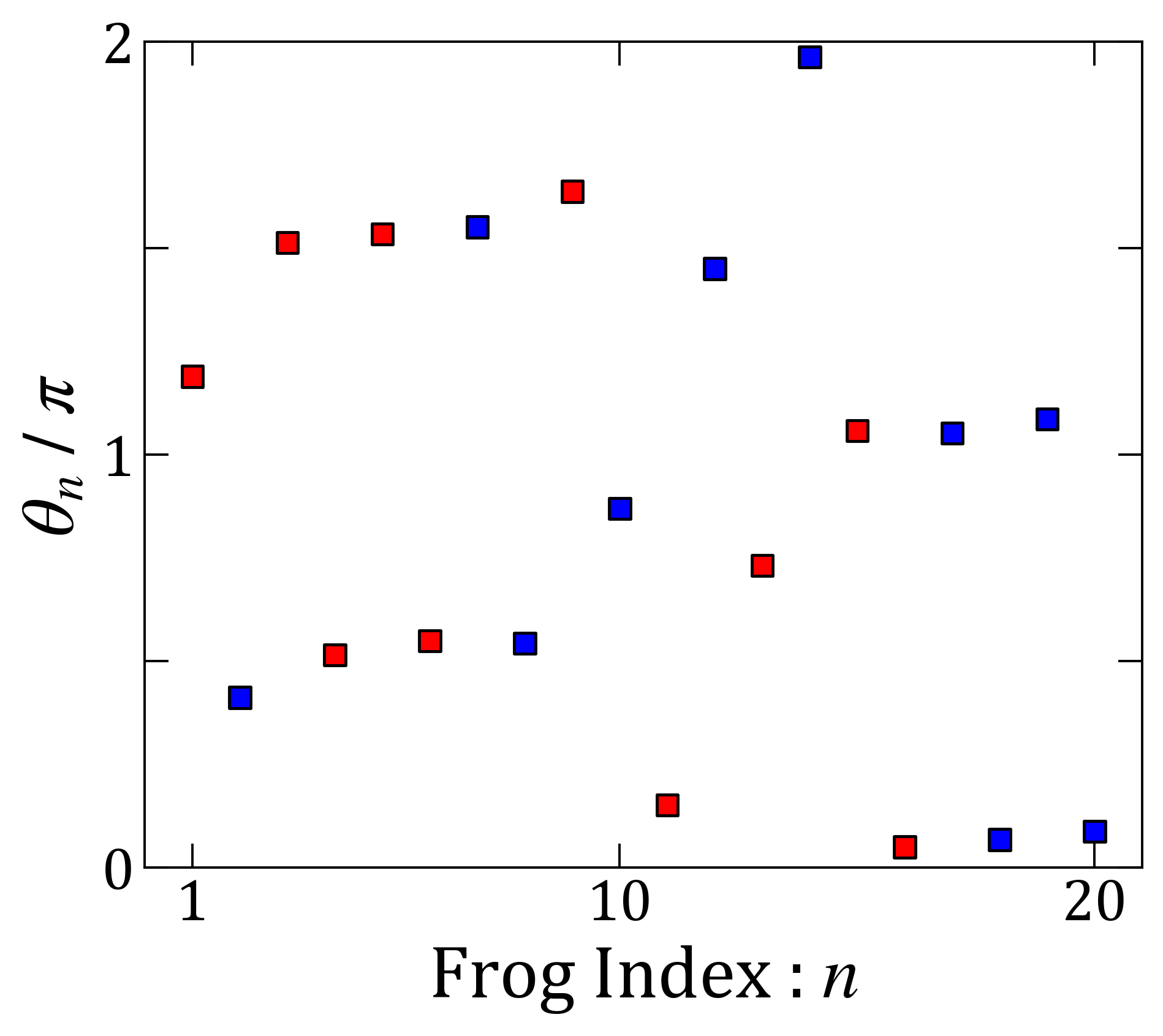}
        \label{fig:KL05_GlobalphaseDist}
    \end{subfigure} \\
    
    \begin{subfigure}[t]{\linewidth}
        \caption{}
        \centering
        \includegraphics[height=3.5cm, keepaspectratio]{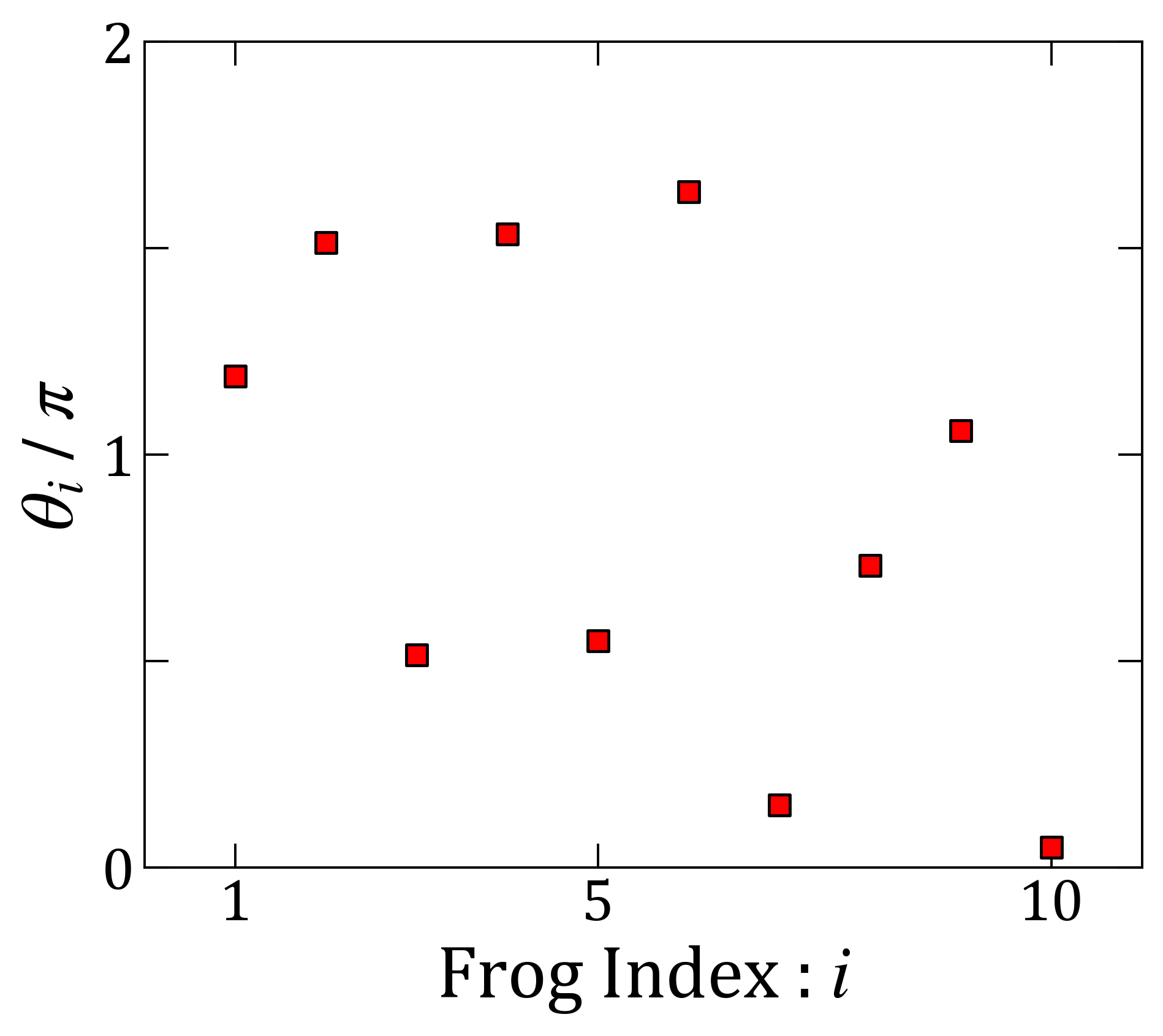}
        \hfill
        \includegraphics[height=3.5cm, keepaspectratio]{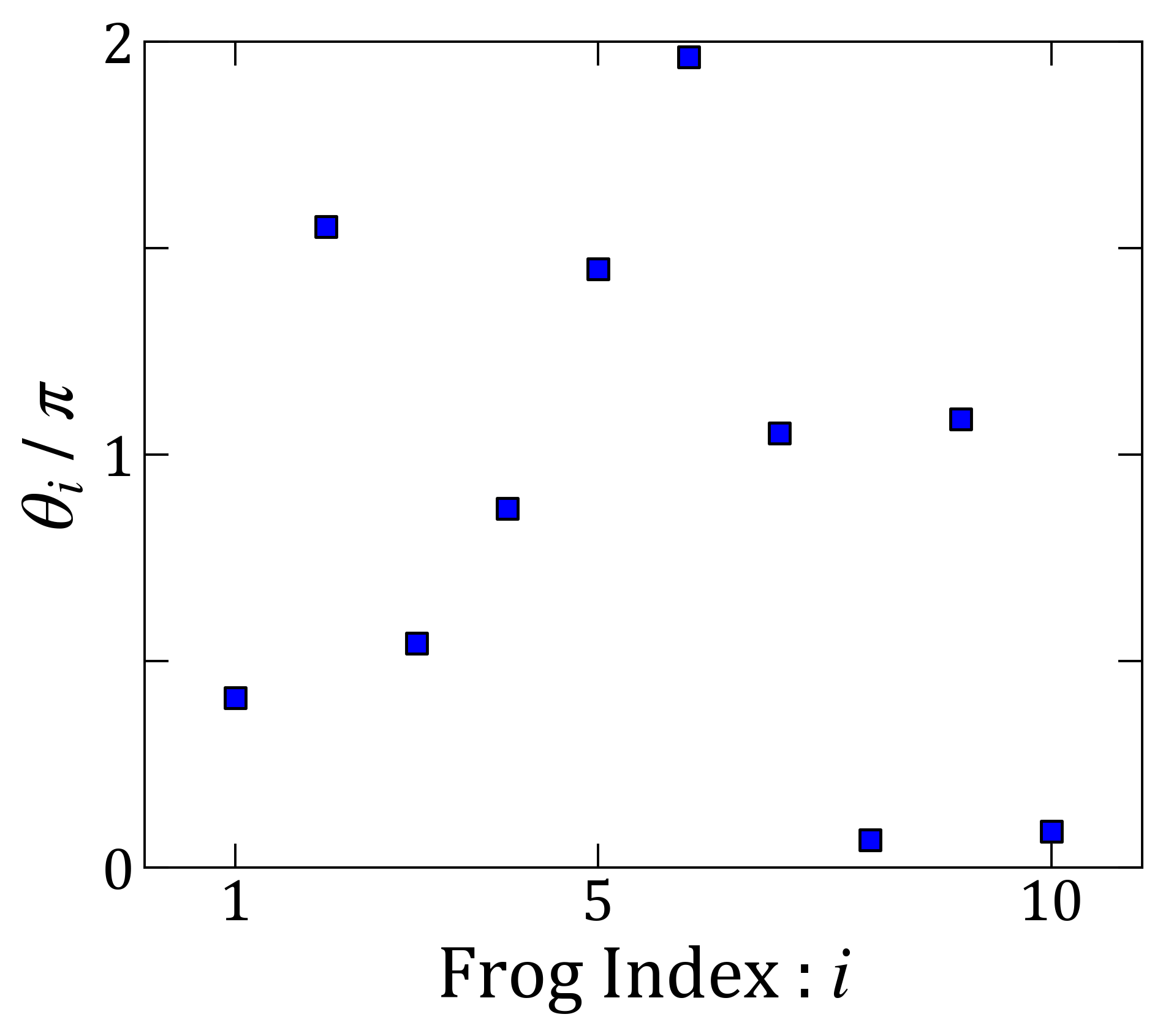}
        \label{fig:KL05_phaseDist}
    \end{subfigure}
    
    \caption{Numerical simulation of our model on the spatio-temporal structure at $\hat{D}_{\mathrm{KL}}=0.5$. 
    (a) The position of male frogs around the breeding site. 
    The males are positioned along the edge of the paddy field at the same inter-frog distance despite of the species
    (b) Anti-phase synchronization with the neighbor of different species. 
    For example, frogs $n=6$ and $7$ synchronize in almost anti-phase, despite they are different species.
    (c) Disruption of two-cluster antisynchronization within the same species.
    The anti-phase synchronization between neighbors in the same species is partially disturbed by the interspecific interaction.}
    \label{figs:KL05_steadyDist}
\end{figure}

Figures \ref{figs:KL1_steadyDist} and \ref{figs:KL05_steadyDist} show how the spatio-temporal structures of frog choruses depend on the value of $\hat{D}_{\mathrm{KL}}$. 
Specifically, we have examined two cases: $\hat{D}_{\mathrm{KL}} = 1$, corresponding to the situation with no interspecific interaction (Fig. \ref{figs:KL1_steadyDist}), 
and $\hat{D}_{\mathrm{KL}} = 0.5$, corresponding to the situation with interspecific interactions whose magnitude is weaker than that of intraspecific interaction (Fig. \ref{figs:KL05_steadyDist}).
For the simulations, the parameter $c_{\mathrm{KL}}$ was set as $c_{\mathrm{KL}}=30$; the same initial condition was used for the two cases. 
Here, we use the index $i$ defined in Section 2.4 to examine the phase difference between neighbors within the same species. 
For $\hat{D}_{\mathrm{KL}} = 0.5$, we also use a different index $n$, which is assigned to all males regardless of species (Fig. \ref{fig:KL05_spatioDist}), in order to examine the phase difference between neighbors across the different species. 
Red and blue plots represent the spatio-temporal structures for each species in which both of the positions and phase difference have converged to equilibrium states.
The simulation with $\hat{D}_{\mathrm{KL}}=1$ demonstrates that (1) the males of each species are positioned along the edge of the field at the same inter-frog distance (Fig. \ref{fig:KL1_spatioDist}) and (2) the pairs of neighboring frogs show anti-phase synchronization, forming the two cluster state \cite{aihara2014_SciRepo} (Fig. \ref{fig:KL1_phaseDist}).
In contrast, the simulation with $\hat{D}_{\mathrm{KL}}=0.5$ demonstrates that (1) the males are positioned along the edge of the field at the almost same inter-frog distance despite of the species (Fig. \ref{fig:KL05_spatioDist}),
(2) the males show anti-phase synchronization not only with neighbors of the same species but also with those of different species (see the phases of Frog IDs $n=6$ and $7$ in Fig. \ref{fig:KL05_GlobalphaseDist}, for example), and (3) the anti-phase synchronization between neighbors in the same species is partially disturbed by the interspecific interaction (Fig. \ref{fig:KL05_phaseDist}).
Thus, the parameter $\hat{D}_{\mathrm{KL}}$, which can be regarded as the similarity of call frequencies, dominantly affects the spatio-temporal structures in the choruses with multiple species.

\begin{figure}[htbp]
    \centering
    \includegraphics[width=0.8\linewidth]{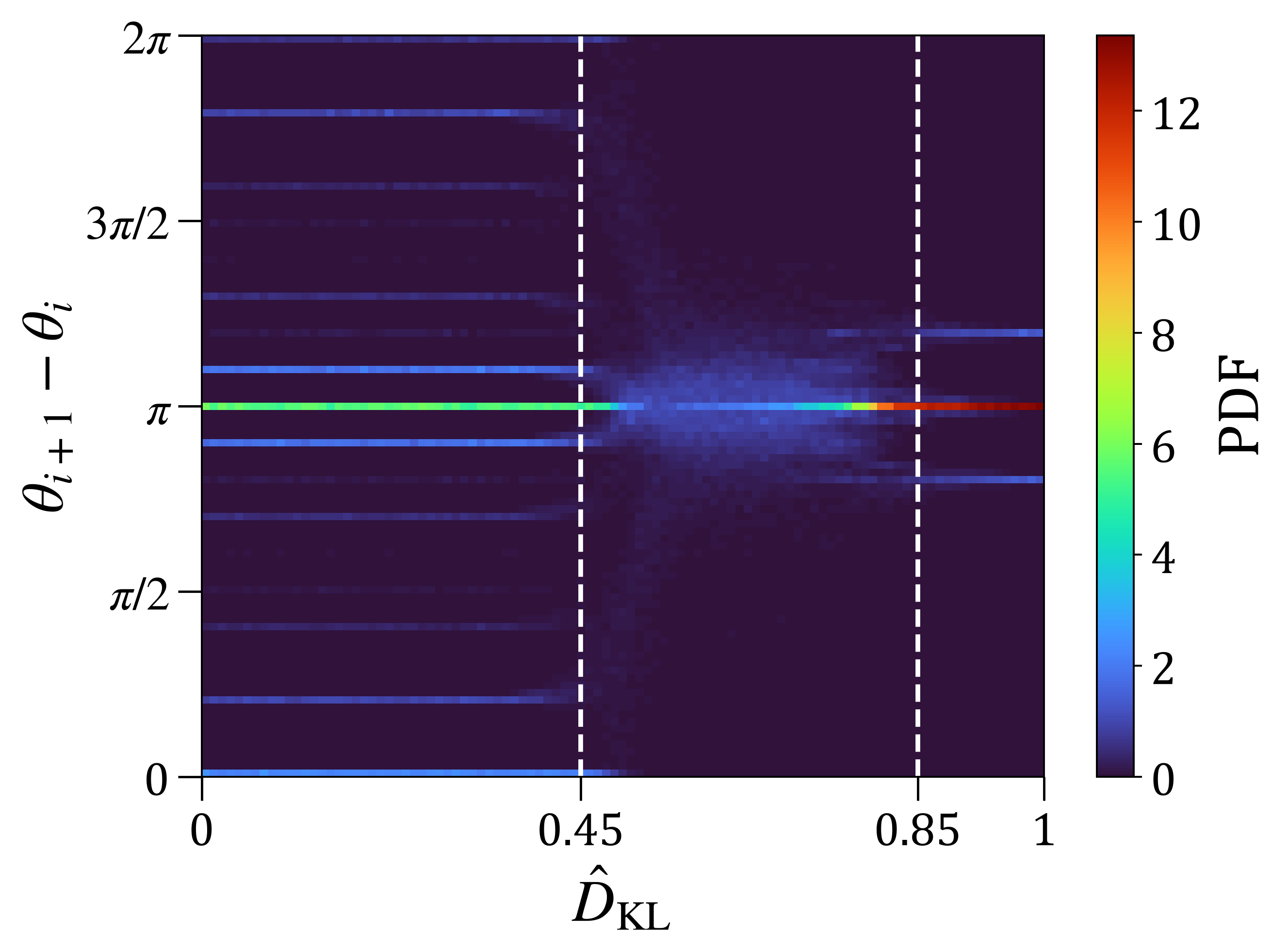}
    \caption{The dependence of phase difference $\theta_{i+1} - \theta_{i}$ on $\hat{D}_{\mathrm{KL}}$. 
    The color represents the probability density. 
    The distributions can be roughly classified into three types based on their shapes: the distribution with multiple peaks ($0 \le \hat{D}_{\mathrm{KL}} < 0.45$), the unimodal distribution with relatively large variance ($0.45 \le \hat{D}_{\mathrm{KL}} < 0.85$), and the distribution with three peaks ($0.85 \le \hat{D}_{\mathrm{KL}}\le 1.00$). 
    }
    \label{fig:phi_DKL_ckl30}
\end{figure}

Figure \ref{fig:phi_DKL_ckl30} shows how the distribution of the phase differences within the same species depends on $\hat{D}_{\mathrm{KL}}$. 
Specifically, we numerically calculated the phase difference between the pair of neighboring frogs in the same species by varying $\hat{D}_{\mathrm{KL}}$ from 0 to 1 in the increments of 0.01. 
At each value of $\hat{D}_{\mathrm{KL}}$, we repeated the simulation at 1,000 times with randomized initial conditions and estimated the probability distribution of the phase differences. 
As shown in Figure \ref{fig:phi_DKL_ckl30}, the shapes of the distributions can be classified into three types based on their shapes. 
First, there are multiple peaks approximately in the range of $0 \le \hat{D}_{\mathrm{KL}} < 0.45$. 
The dominant peak around $\theta_{i+1} - \theta_{i}=\pi$ corresponds to the two-cluster antisynchronization while the relatively small peaks, such as $\theta_{i+1} - \theta_{i}=0.9\pi,~ 1.1\pi$, correspond to the wavy antisynchronization (see Appendix C for details) that has been reported in \cite{aihara2014_SciRepo}.
Second, there is an unimodal and symmetric peak around $\theta_{i+1} - \theta_{i}=\pi$ with relatively larger variance approximately in the range of $0.45 \le \hat{D}_{\mathrm{KL}} < 0.85$.
This peak becomes sharper as $\hat{D}_{\mathrm{KL}}$ increases. 
Third, there are three peaks approximately in the range of $0.85 \le \hat{D}_{\mathrm{KL}} \le 1.00$.
The dominant peak around $\theta_{i+1} - \theta_{i}=\pi$ corresponds to the two-cluster antisynchronization while the remaining peaks around $\theta_{i+1} - \theta_{i}=0.8\pi$ or $1.2\pi$ correspond to wavy antisynchronization. 

\subsection{Quantification of the spatio-temporal strutucre using the order parameter}

\begin{figure}[htbp]
    \centering
    \begin{subfigure}[b]{0.7\linewidth}
        \caption{}
        \centering
        \includegraphics[width=\linewidth]{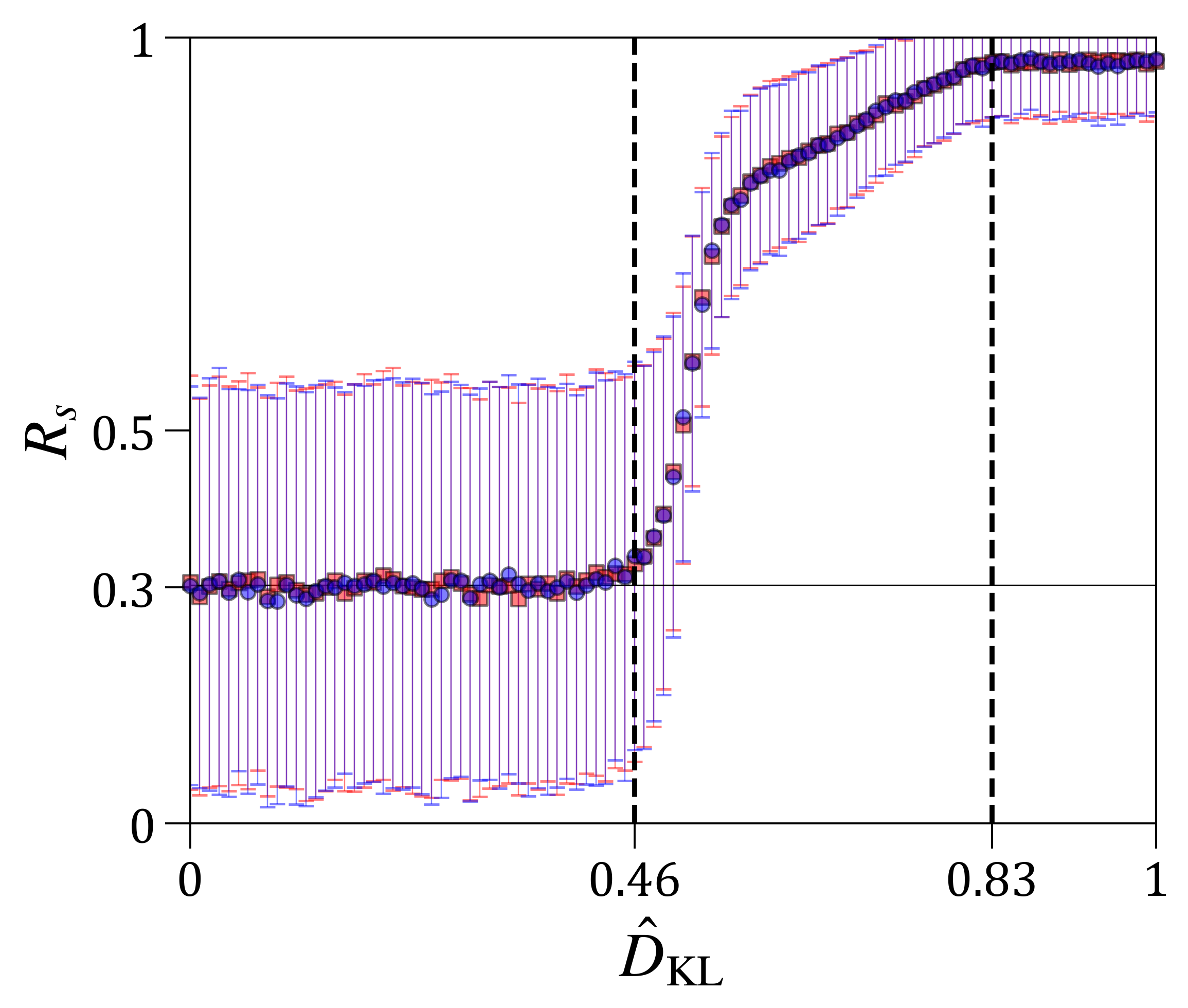}
        \label{fig:ckl30_df}
    \end{subfigure}\\
    \begin{subfigure}[b]{0.7\linewidth}
        \caption{}
        \centering
        \includegraphics[width=\linewidth]{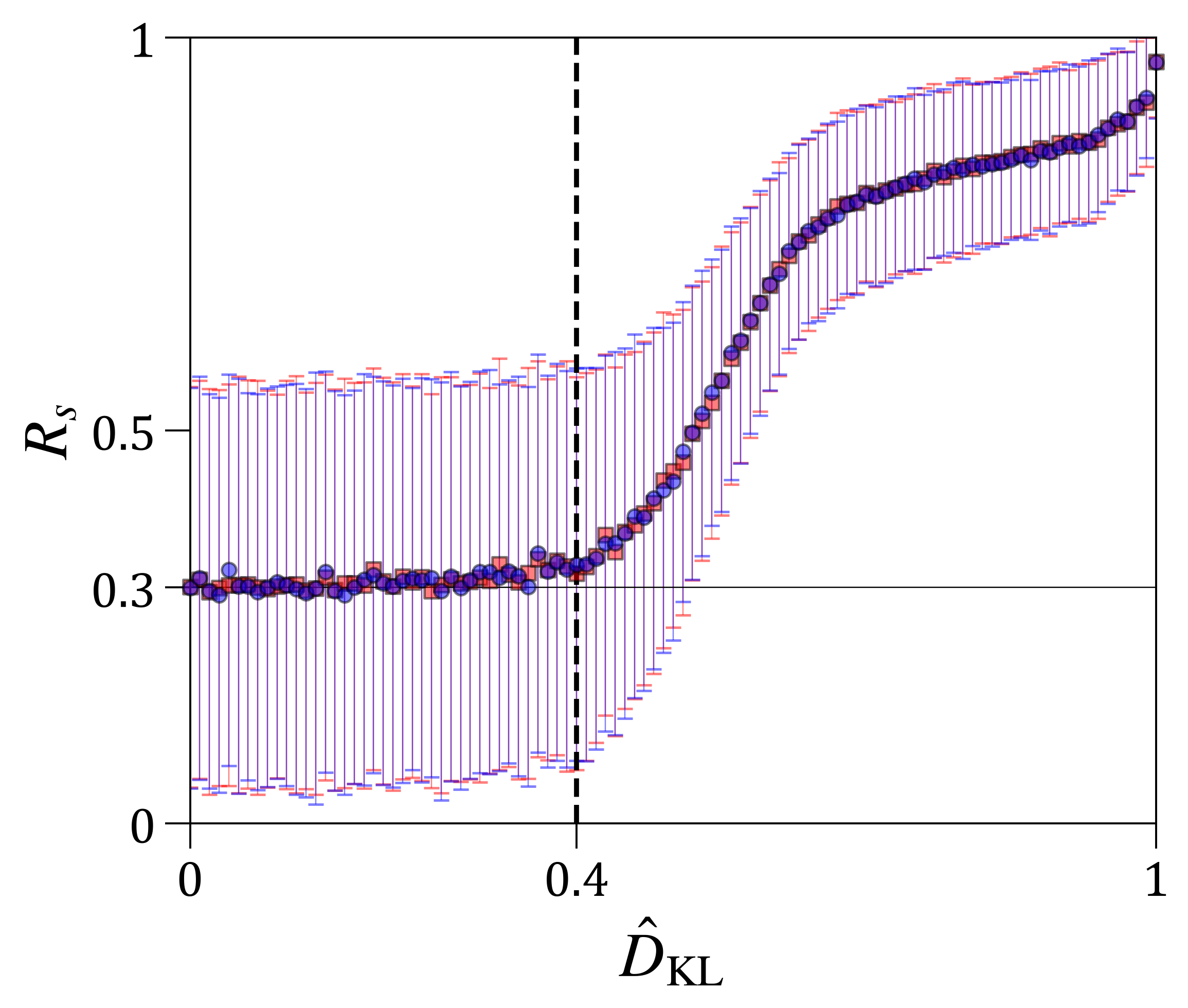}
        \label{fig:ckl10_df}
    \end{subfigure}
    
    \caption{The dependence of the order parameter $R_s$ on $\hat{D}_{\mathrm{KL}}$ for the two cases: (a) $c_{\mathrm{KL}}=30$ and (b) $c_{\mathrm{KL}}=10$. 
    Each dot and bar represent the mean and the standard deviation of the order parameter $R_s$, respectively. 
    In the case of $c_{\mathrm{KL}}=30$, the mean of the order parameter $R_s$ remains the constant lower value around 0.3 in the range of $0 \le \hat{D}_{\mathrm{KL}} < 0.46$, rapidly increases towards the maximum value 1.0 in the range of $0.46 \le \hat{D}_{\mathrm{KL}} < 0.83$, and  remains around 1.0 in the range of $0.83 \le \hat{D}_{\mathrm{KL}} \le 1.00$.
    In the case of $c_{\mathrm{KL}}=10$, the mean of the order parameter $R_s$ remains the constant lower value around 0.3 in the range of $0 \le \hat{D}_{\mathrm{KL}} < 0.40$ and rapidly increases towards the maximum value 1.0 in the range of $0.40 \le \hat{D}_{\mathrm{KL}} \le 1.00$.}
    \label{figs:ckl_Rs_df}
\end{figure}

Figure \ref{figs:ckl_Rs_df} shows how the order parameter $R_s$ depends on $\hat{D}_{\mathrm{KL}}$ when the spatio-temporal structure of our model has converged to an equilibrium state. 
The value of $\hat{D}_{\mathrm{KL}}$ was varied from $0$ to $1$ in the increments of $0.01$.
Given the dependence of $R_s$ on initial condition, we performed numerical simulations at 1,000 times for each $\hat{D}_{\mathrm{KL}}$ with randomized initial conditions. 
Here, we examined the two cases of $c_{\mathrm{KL}}=30$ and $c_{\mathrm{KL}}=10$ to vary the shape of the logistic function in Equation \ref{K_mn}. 
When $c_{\mathrm{KL}}=30$, the dependence of $R_s$ on $\hat{D}_{\mathrm{KL}}$ can be categorized into three types (Fig. \ref{fig:ckl30_df}). 
First, $R_s$ takes the consistent value around 0.30 approximately in the range of $0 \le \hat{D}_{\mathrm{KL}} < 0.46$.
The value $R_s \approx 0.30$ can be theoretically estimated as the lower bound of the expected value of the order parameter (see Appendix D for details). 
Second, $R_s$ shows a rapid rise with the decrease of its standard deviation approximately in the range of $0.46 \le \hat{D}_{\mathrm{KL}} < 0.83$ as $\hat{D}_{\mathrm{KL}}$ increases. 
Third, $R_s$ takes the almost maximum value of 1.00 approximately in the range of $0.83 \le \hat{D}_{\mathrm{KL}} \le 1.00$.
The above feature is consistent with the probability distributions of $\theta_{i+1}-\theta_i$ (Fig. \ref{fig:ckl30_df}) whose structure can be also categorized into three types.  
In contrast, the dependence of $R_s$ on $\hat{D}_{\mathrm{KL}}$ can be categorized into two types when $c_{\mathrm{KL}}=10$ (Fig. \ref{fig:ckl10_df}). 
The simulation demonstrates that (1) $R_s$ takes the consistent value around 0.30 in the range of $0 \le \hat{D}_{\mathrm{KL}} < 0.40$ and (2) $R_s$ shows a rapid rise with the decrease of its standard deviation in the range of $0.40 \le \hat{D}_{\mathrm{KL}} \le 1.00$ as $\hat{D}_{\mathrm{KL}}$ increases, which is consistent with the case of $c_{\mathrm{KL}}=30$. 
As for the latter result, the mean of $R_s$ reaches the almost maximum value of 1 at $\hat{D}_{\mathrm{KL}} = 1$.
However, the region in which the mean of $R_s$ remain around 1 (corresponding to the third region in Fig. \ref{fig:ckl30_df}) has disappeared in this case of $c_{\mathrm{KL}}=10$. 
These results have indicated that (1) the two-cluster antisynchronization can be established when the distributions of call frequencies are much different, corresponding to $R_s \approx 1.0$ with the larger $\hat{D}_{\mathrm{KL}}$, 
and (2) the two-cluster antisynchronization can be disturbed due to the interspecific interaction when the distribution of call frequencies are similar, corresponding to $R_s \approx 0.3$ with the smaller $\hat{D}_{\mathrm{KL}}$.

\section{DISCUSSION}
In this study, we investigated the spatio-temporal structures of frog choruses with two species both empirically and theoretically. 
First, we performed a playback experiment on male Japanese treefrogs and observed anti-phase synchronization in response to calls of other acoustic animals. 
The analysis of audio recordings indicated that the degree of anti-phase synchronization was affected by the similarity of call frequencies.
Second, we proposed a mathematical model on the basis of the empirical data and carried out numerical simulations by varying the similarity of call frequencies via the normalized KL divergence.
The simulation showed that (1) the two-cluster antisynchronization is established when the distributions of call frequencies are much different between two species, and (2) the two-cluster antisynchronization is disturbed due to the interspecific interaction when the distribution of call frequencies are similar.

Here we discuss the relationship between the simulation and empirical data.
Playback experiments demonstrated that male Japanese tree frogs synchronized in anti-phase with the sound stimuli of \textit{F. kawamurai} (Fig. \ref{figs_RealFrog:calling behaivior}(\subref{fig_RealFrog:numa2ama_histgram})) but did not synchronize with the stimuli of \textit{M. siamensis} (Fig. \ref{figs_RealFrog:calling behaivior}(\subref{fig_RealFrog:tambo2ama_histgram})).
The difference in call frequencies between Japanese tree frogs and \textit{F. kawamurai} was estimated as 2,100 Hz while that between the tree frogs and \textit{M. siamensis} was estimated as 3,400 Hz.
According to the procedure using the KL divergence (Section 2.3), the difference of 2,100 Hz was quantified as $\hat{D}_{\mathrm{KL}} \approx 0.38$ while the difference of 3,400 Hz was quantified as $\hat{D}_{\mathrm{KL}}=1.00$ (Section 2.3). 
Then, numerical simulations of our model revealed the dependence of the two-cluster antisynchronization on the similarity of call frequencies. 
Specifically, the simulations with $c_{KL}=30$ and $10$ showed that the mean of $R_s$ took the minimum value of 0.30 at $\hat{D}_{\mathrm{KL}}=0.38$ while it took the large value around 1.00 at $\hat{D}_{\mathrm{KL}}=1.00$ (Figs. \ref{fig:ckl30_df} and \ref{fig:ckl10_df}).
This result predicts that the calls of \textit{F. kawamurai} likely disturb the anti-phase synchronization of Japanese tree frogs even in natural environments while the calls of \textit{M. siamensis} do not.
Recently, we have succeeded in quantifying the spatio-temporal frequency features of frog choruses in natural environment by using the microphone arrays and sound-imaging devices \cite{aihara2026arXiv}; this methodology is applicable to the validation of the numerical prediction based on empirical data. 
Moreover, the combination of the field recordings with the methodology of model identification, such as those proposed by Ota and Aoyagi (2014) \cite{ota2014direct} and Mori and Kori (2022) \cite{mori2022noninvasive}, likely allows us to examine the behavioral mechanism of frog choruses for details in the context of the precise modeling. 

Next, we address the contribution and future issue of the proposed model from the viewpoint of spatio-temporal dynamics in a system of coupled mobile oscillators (e.g., swarmalators \cite{okeeffe2017swarmalators}). 
In this study, we model the choruses of male frogs with two species by using Equations (\ref{phase dynamics})--(\ref{K_mn}) in which two kinds of repulsive interactions, i.e., intraspecific repulsive interaction and interspecific repulsive interaction, are assumed. 
While the previous studies often focus on a purely attractive interaction or mixed interactions with attractive and repulsive effects \cite{daido1992quasientrainment, hong2011conformists, abrams2008solvable, ghosh2023antiphaseSwarmalators}, the systems with two kinds of repulsive interactions remain relatively unexplored. 
This study has demonstrated that the order parameter $R_s$ quantifying the occurrence of anti-phase synchronization within the same species keeps the lower value and then rapidly increases as $\hat{D}_{\mathrm{KL}}$ increases.
Such a transition of the order parameter in a system of coupled mobile oscillators with two kinds of repulsive interactions has been newly reported in this study.
Furthermore, we have observed that anti-phase synchronization between neighbors can be disturbed in a local population especially in the medium value of $\hat{D}_{\mathrm{KL}}$ (see Fig. \ref{fig:KL05_phaseDist} for the case of $\hat{D}_{\mathrm{KL}}=0.5$).
It remains a future issue to examine the mechanism of this phenomenon from the viewpoint of chimera state in a system of coupled oscillators \cite{kuramoto2002coexistence}. 

Here, we discuss the results of the playback experiment from a biological point of view. 
As shown in Figures \ref{fig_RealFrog:numa2ama_histgram} and \ref{fig_RealFrog:tambo2ama_histgram}, Japanese tree frogs tend to synchronize in anti-phase with sound stimuli of \textit{F. kawamurai} but do not synchronize with the stimuli of \textit{M. siamensis}.
The point is that the call frequency of Japanese tree frogs is similar to that of  \textit{F. kawamurai} but is much different from that of \textit{M. siamensis}.
Given that the frogs produce sounds to attract conspecific females and also advertise their territories to competitors \cite{gerhardt2002acoustic}, Japanese tree frogs likely attempt to synchronize in anti-phase with \textit{F. kawamurai} and reduce the acoustic interference originating from the similar call frequencies.
It should be noted that the histogram of phase difference obtained from the experiment with the stimuli of \textit{F. kawamurai} shows not only the dominant peak around $\phi_{AB} = \pi$, but also the smaller peak around $\phi_{AB} = 5/3 \pi$. 
The biological meaning of this small peak remains unclear and needs further examination.

\section*{Appendix A: Playback System}
We developed a real-time playback system for our experiments to control the playback of pre-recorded frog calls and analyze the responses of real frogs.
The onset times of real frog calls are required to analyze the interactions between the frogs and the playback system. 
We detect these onset times by applying audio signal processing techniques implemented in a real-time system written in Python.

\begin{figure}[t]
  \centerline{\includegraphics[width=0.99\columnwidth]{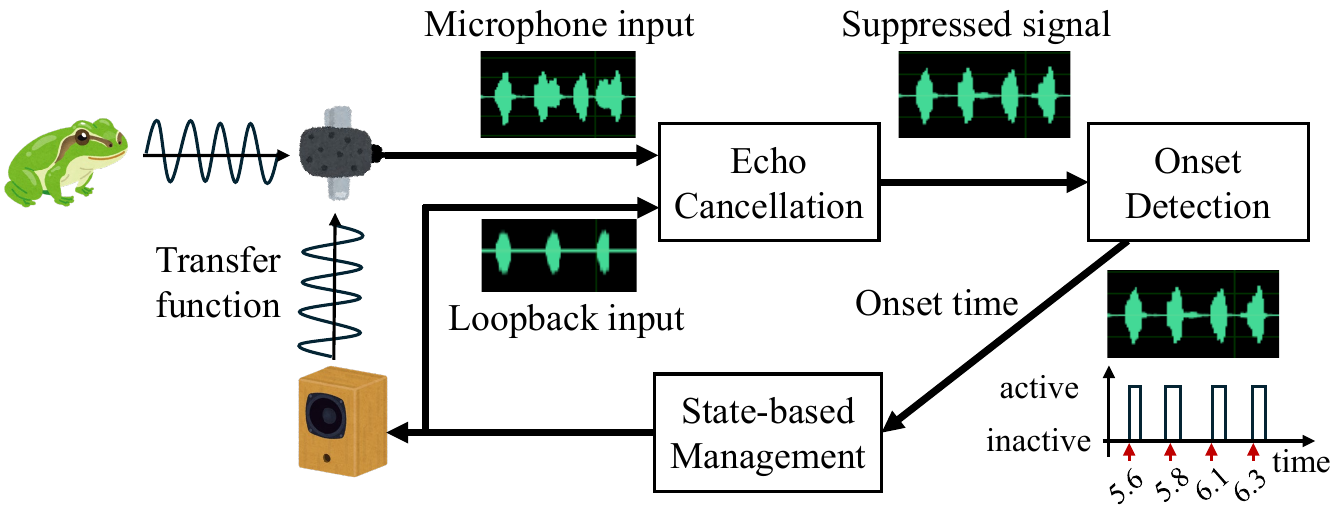}}
  \caption{Flow of our playback system}
  \label{fig:system}
\end{figure}

Our playback system mainly consists of three modules: echo cancellation, onset detection, and state-based management. 
The processing flow is illustrated in Fig.\ref{fig:system}.

\paragraph{Echo Cancellation}
The echo cancellation module suppresses loopback signals contained in the microphone input because the microphone captures a mixture of real frog calls and sounds reproduced by the loudspeaker.
First, the loopback signal observed at the microphone is estimated by applying a transfer function from the loudspeaker to the microphone to the playback signal. 
The estimated signal is then subtracted from the microphone input to extract the real frog calls. 
In this experiment, the transfer function was measured in advance.

\paragraph{Onset Detection}

The onset detection module estimates the onset times of real frog calls from the echo-suppressed signal. 
First, the signal power is calculated for each time frame, and each frame is classified as either active or inactive by thresholding the signal power.
An online smoothing method is then applied to the frame-level activity sequence to reduce fluctuations. 
The onset time is estimated by detecting transitions from an inactive to an active state.

\paragraph{State-based Management}
The state-based management module determines when pre-recorded frog calls should be played through the loudspeaker and controls their playback timing.
The system has two main states: an initial state and a reaction state. 
In the initial state, the system plays a pre-recorded frog call and transitions to the reaction state after playback is completed.
In the reaction state, the system plays another pre-recorded frog call when ten call onsets are detected. 
If no call onset is detected for a predefined period, the system returns to the initial state and plays the first pre-recorded frog call.

\section*{Appendix B: The difference in standard deviations of call frequency}

Here, we explain how to estimate the standard deviations of the distributions of call frequency around the dominant peaks using empirical data of Japanese tree frog, \textit{F. kawamurai}, and \textit{M. siamensis}.

\begin{figure}
    \centering
    \includegraphics[width=0.7\linewidth]{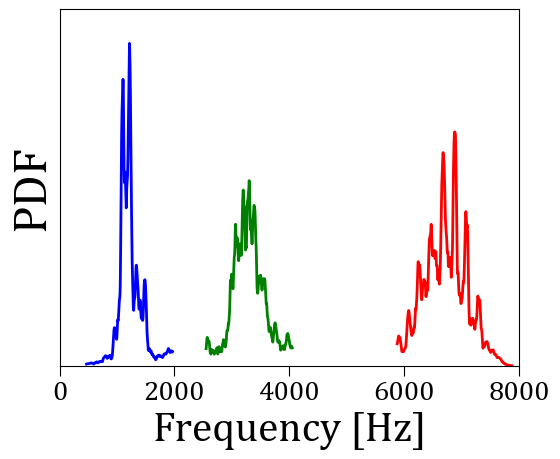}
    \caption{The extracted power spectra of call frequencies for three species of acoustic animals.
    The green, blue, and red lines represent the distributions of call frequency around the dominant peaks for Japanese tree frog, \textit{F. kawamurai}, and \textit{M. siamensis}, respectively.} 
    \label{fig:call freq. around the dominant peaks}
\end{figure}

In this study, we model the distribution around the dominant peak as a normal distribution (see Section 2.3). 
To fit the data with such a unimodal distribution, we have first extracted the power spectrum in a specific range that contains the dominant peak but do not include other peaks. 
Accordingly, the range was set to $\pm 750$ Hz for the dominant peaks of Japanese tree frog and \textit{F. kawamurai}, whereas it was set to $\pm 1{,}000$ Hz for the dominant peak of \textit{M. siamensis} (Fig. \ref{fig:call freq. around the dominant peaks}). 
Then, we have calculated the standard deviation for each spectrogram. 
It has been shown that (1) the standard deviation of Japanese tree frog is 288 Hz, (2) the standard deviation of \textit{F. kawamurai} is 209 Hz, and (3) the standard deviation of \textit{M. siamensis} is 365 Hz.
So the standard deviation of \textit{F. kawamurai} is approximately 0.7 times that of Japanese tree frog, whereas that of \textit{M. siamensis} is approximately 1.3 times that of Japanese tree frog. 
Based on these results, we have approximated the standard deviations of the three species to be the same (see Section 2.3 of the main manuscript).

\section*{Appendix C: Spatio-temporal structure for $D_{\mathrm{KL}}=0$}

Here, we explain the detailed spatio-temporal structures for $\hat{D}_{\mathrm{KL}}=0$. 
The point is that $\hat{D}_{\mathrm{KL}}=0$ corresponds to the situation in which the distributions of call frequencies for the two species are identical (Fig. \ref{fig: normal distributions} in the main manuscript), inducing the same magnitude in intraspecific interaction and interspecific interaction.
Subsequently, this assumption of $\hat{D}_{\mathrm{KL}}=0$ indicates that (1) the two species of male frogs cannot discriminate between each other and (2) the two species behave like a single species. 
To quantify the chorus structure across the two species, we use the index $n$, which is assigned to all males. 
Then, we also use the index $i$, which is assigned only to males of each species, so as to examine the structure within the same species.

\begin{figure}[htbp]
    \centering
    \begin{subfigure}[t]{0.48\linewidth}
        \vspace{0pt}
        \caption{}
        \centering
        \includegraphics[height=3.5cm, keepaspectratio]{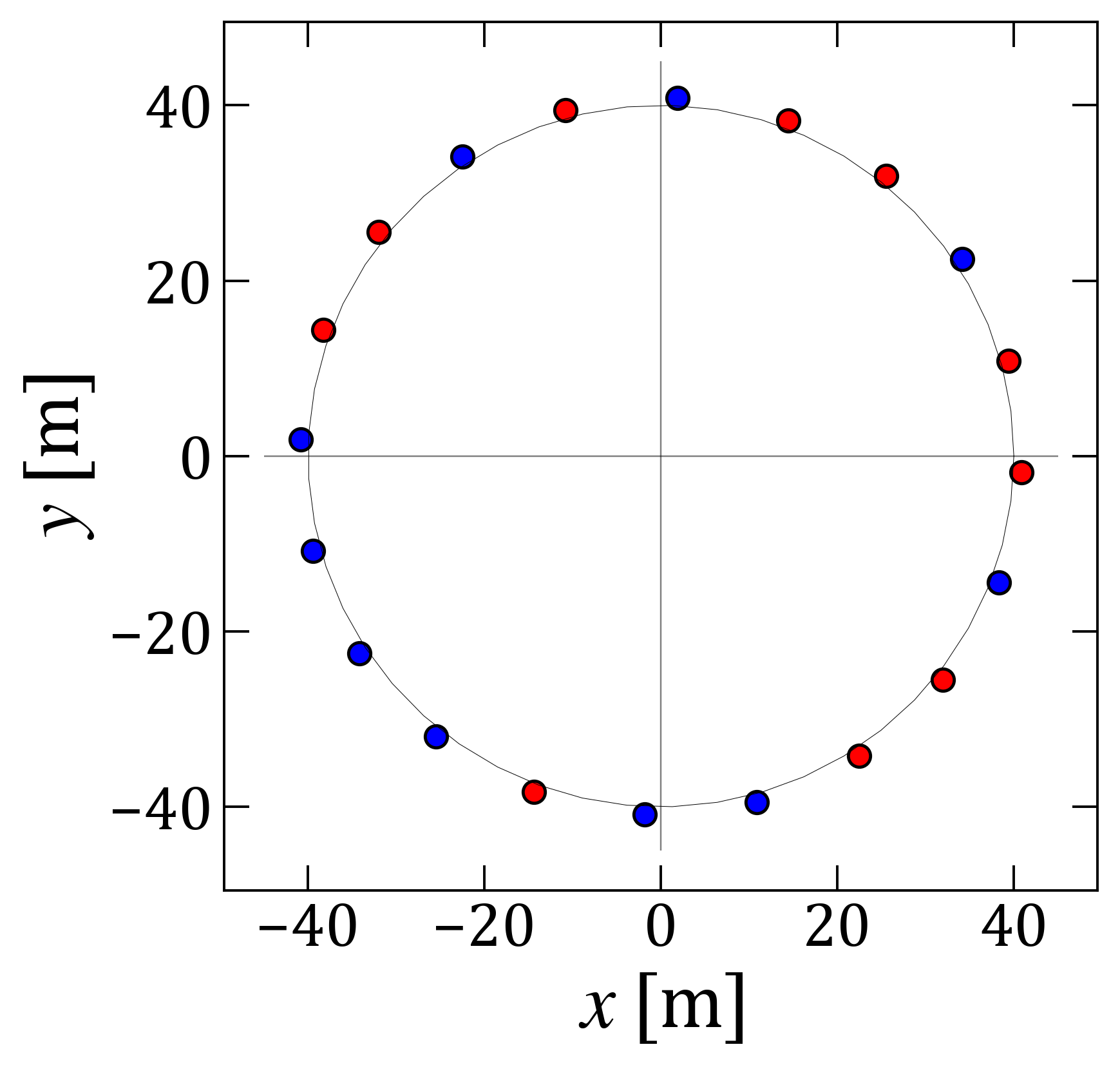}
        \label{fig:KL0_spatioDist_2cluster}
    \end{subfigure}
    \begin{subfigure}[t]{0.48\linewidth}
        \vspace{0pt}
        \caption{}
        \centering
        \includegraphics[height=3.5cm, keepaspectratio]{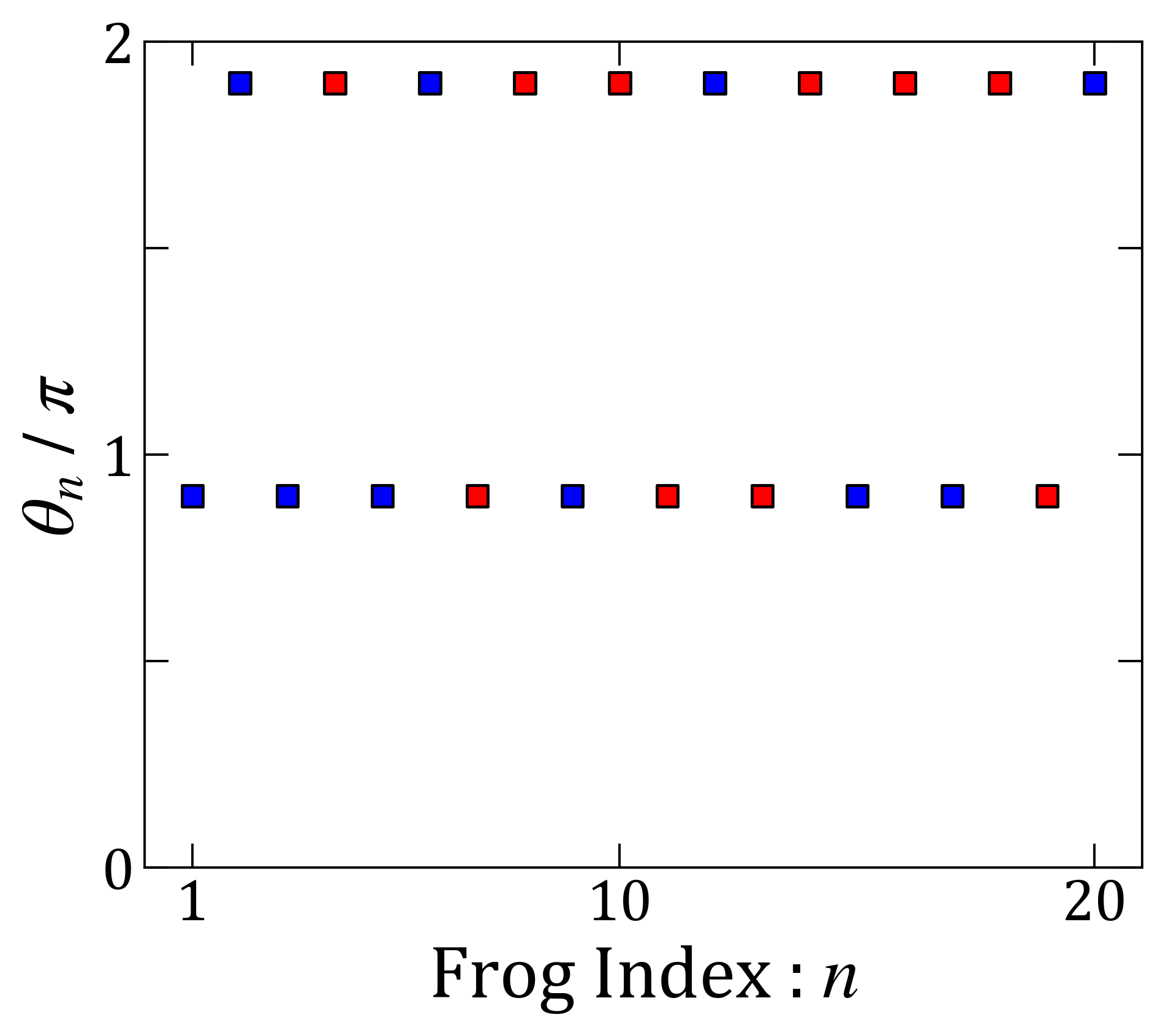}
        \label{fig:KL0_GlobalphaseDist_2cluster}
    \end{subfigure}\\

    \begin{subfigure}[t]{\linewidth}
        \vspace{0pt}
        \caption{}
        \centering
        \includegraphics[height=3.5cm, keepaspectratio]{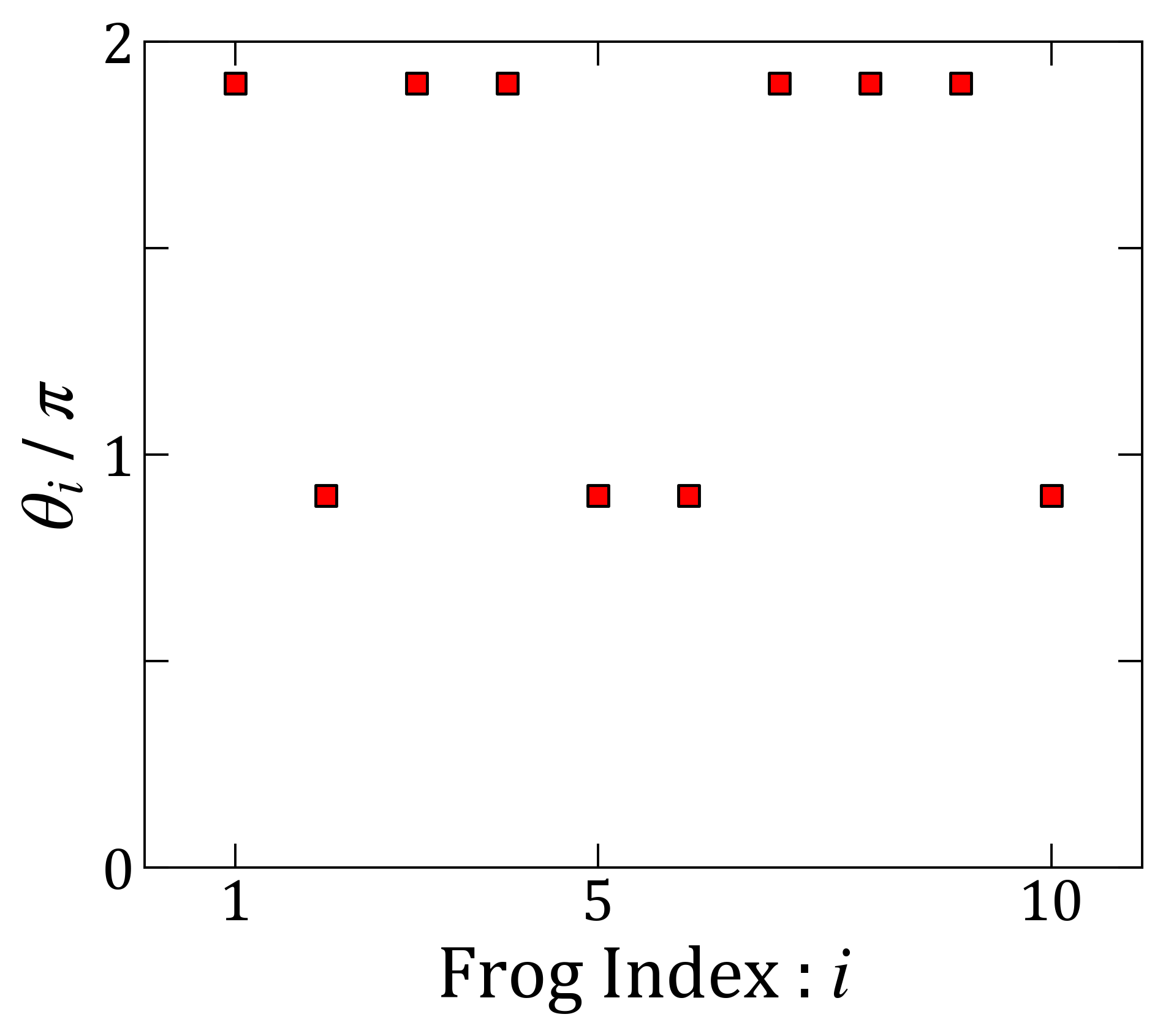}
        \hfill
        \includegraphics[height=3.5cm, keepaspectratio]{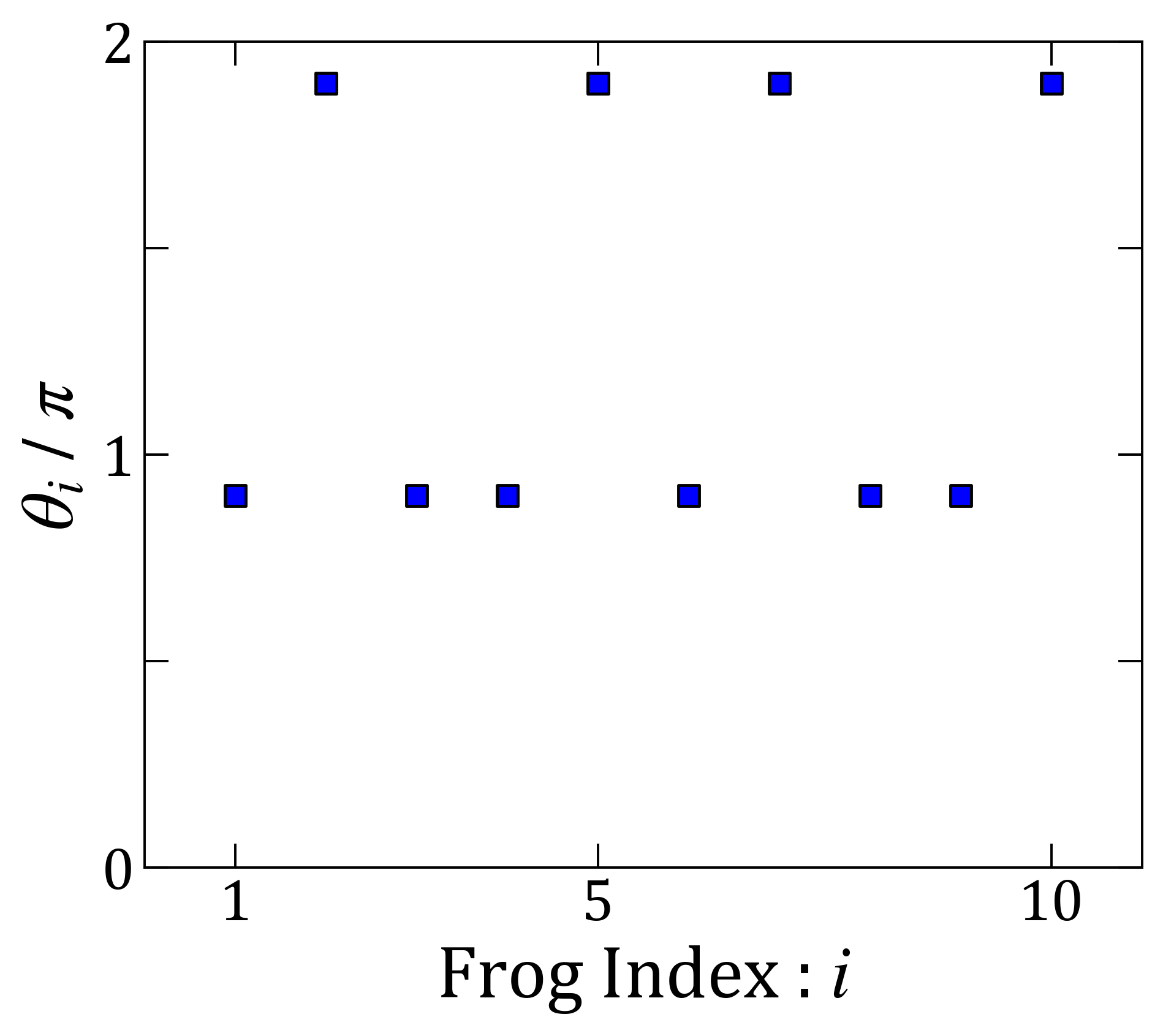}   
        \label{fig:KL0_phaseDist_2cluster}        
    \end{subfigure}

    \caption{Two-cluster antisynchronization obtained from numerical simulation at $\hat{D}_{\mathrm{KL}}=0$. 
    (a) Spatial structure in a frog chorus. The males are positioned along the edge of the field at the same inter-frog distance despite of the species. 
    (b) Two-cluster antisynchronization in a frog chorus. 
    Here the index $n$ is assigned to all males of the two species. 
    The males synchronize in anti-phase not only with neighbors of same species, but also with those of a different species.
    (c) Mixture of anti-phase and in-phase synchronization in the same species. 
    The males synchronize not only in anti-phase but also in-phase with neighbors of the same species. }
    \label{fig:Two-cluster; D_KL=0}
\end{figure}

\begin{figure}[htbp]
    \centering
    \begin{subfigure}[t]{0.48\linewidth}
        \vspace{0pt}
        \caption{}
        \centering
        \includegraphics[height=3.5cm, keepaspectratio]{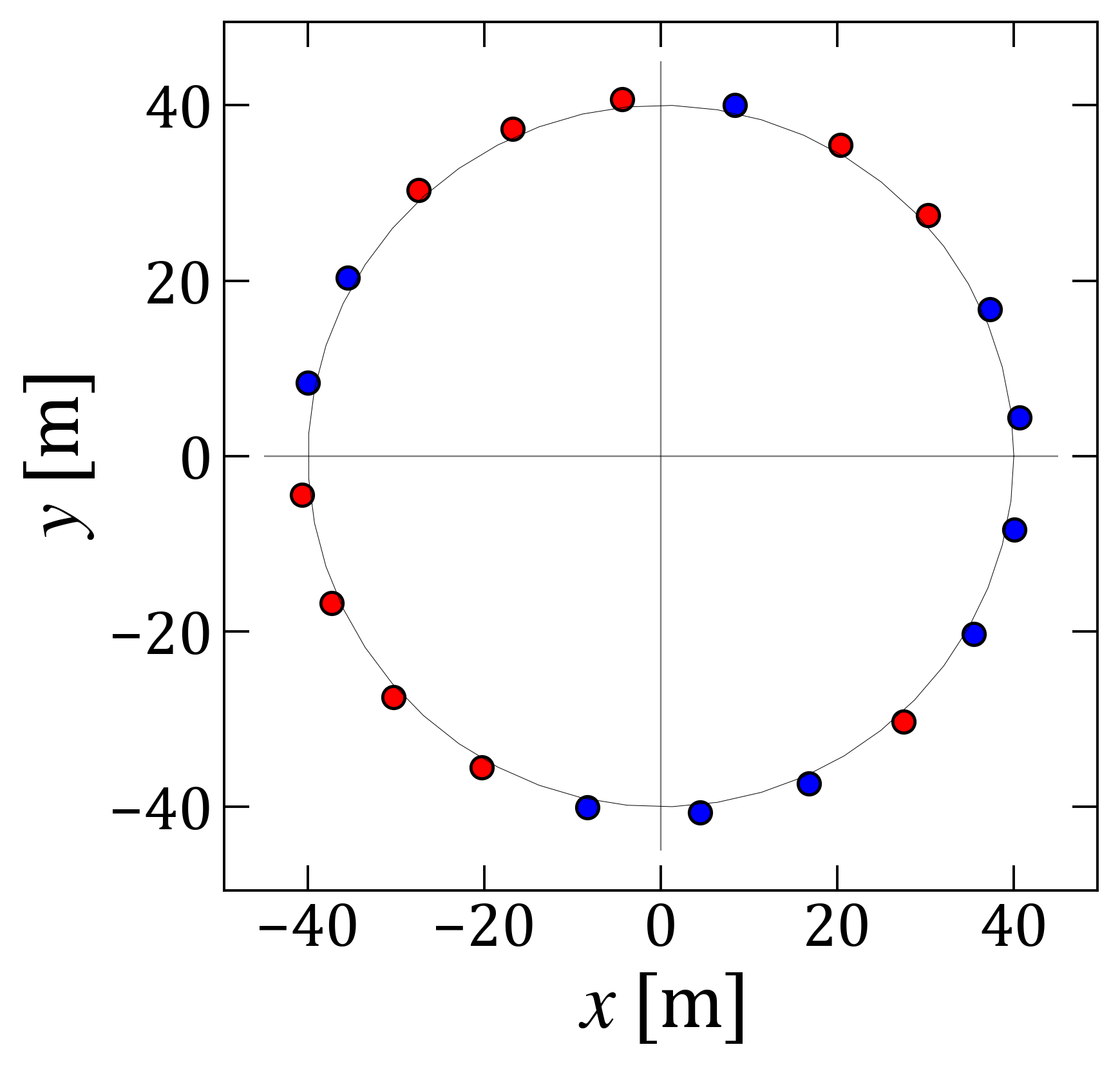}
        \label{fig:KL0_spatioDist_wavy}
    \end{subfigure}
    \begin{subfigure}[t]{0.48\linewidth}
        \vspace{0pt}
        \caption{}
        \centering
        \includegraphics[height=3.5cm, keepaspectratio]{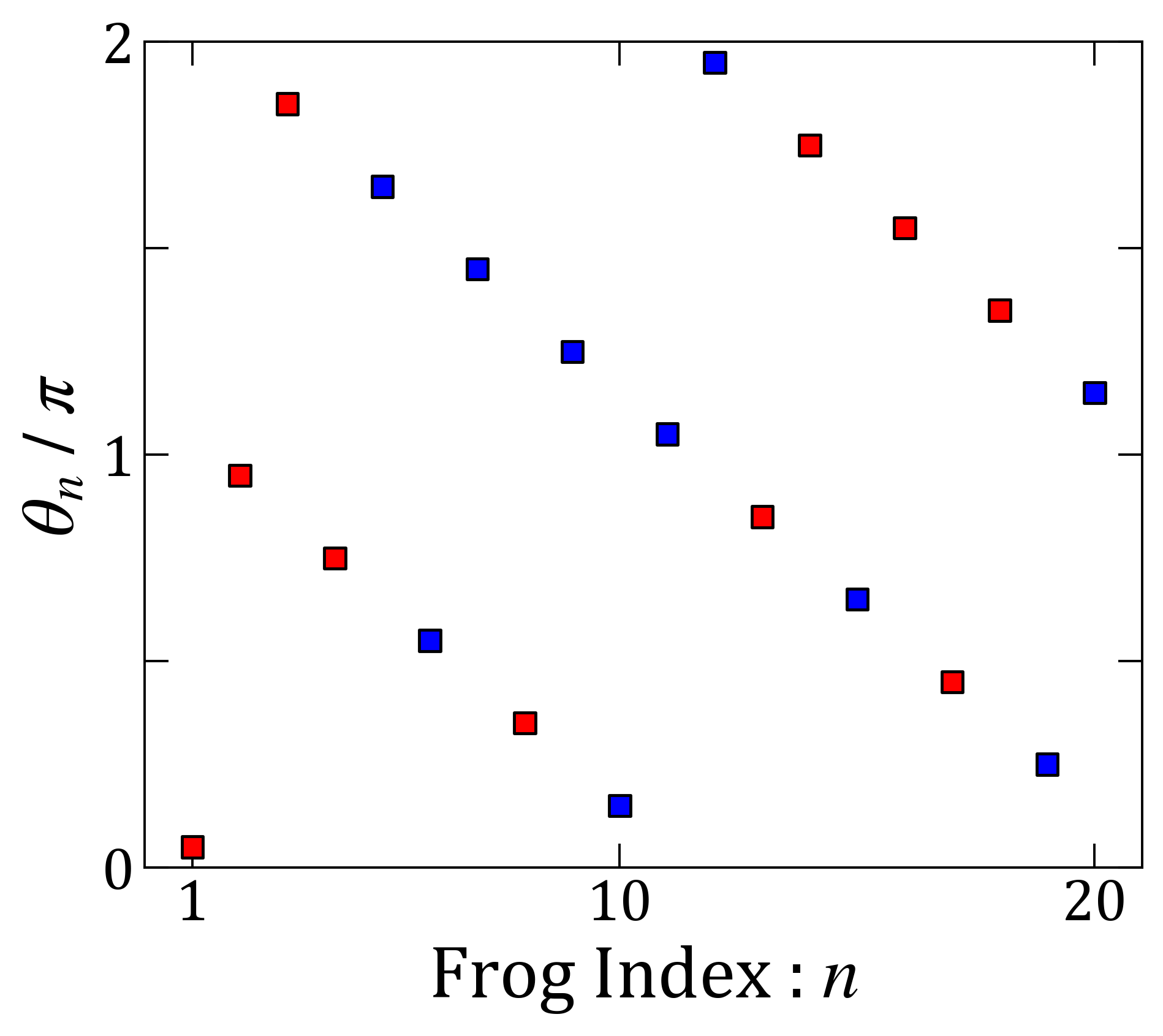}
        \label{fig:KL0_GlobalphaseDist_wavy}
    \end{subfigure}

    \begin{subfigure}[t]{\linewidth}
        \vspace{0pt}
        \caption{}
        \centering
        \includegraphics[height=3.5cm, keepaspectratio]{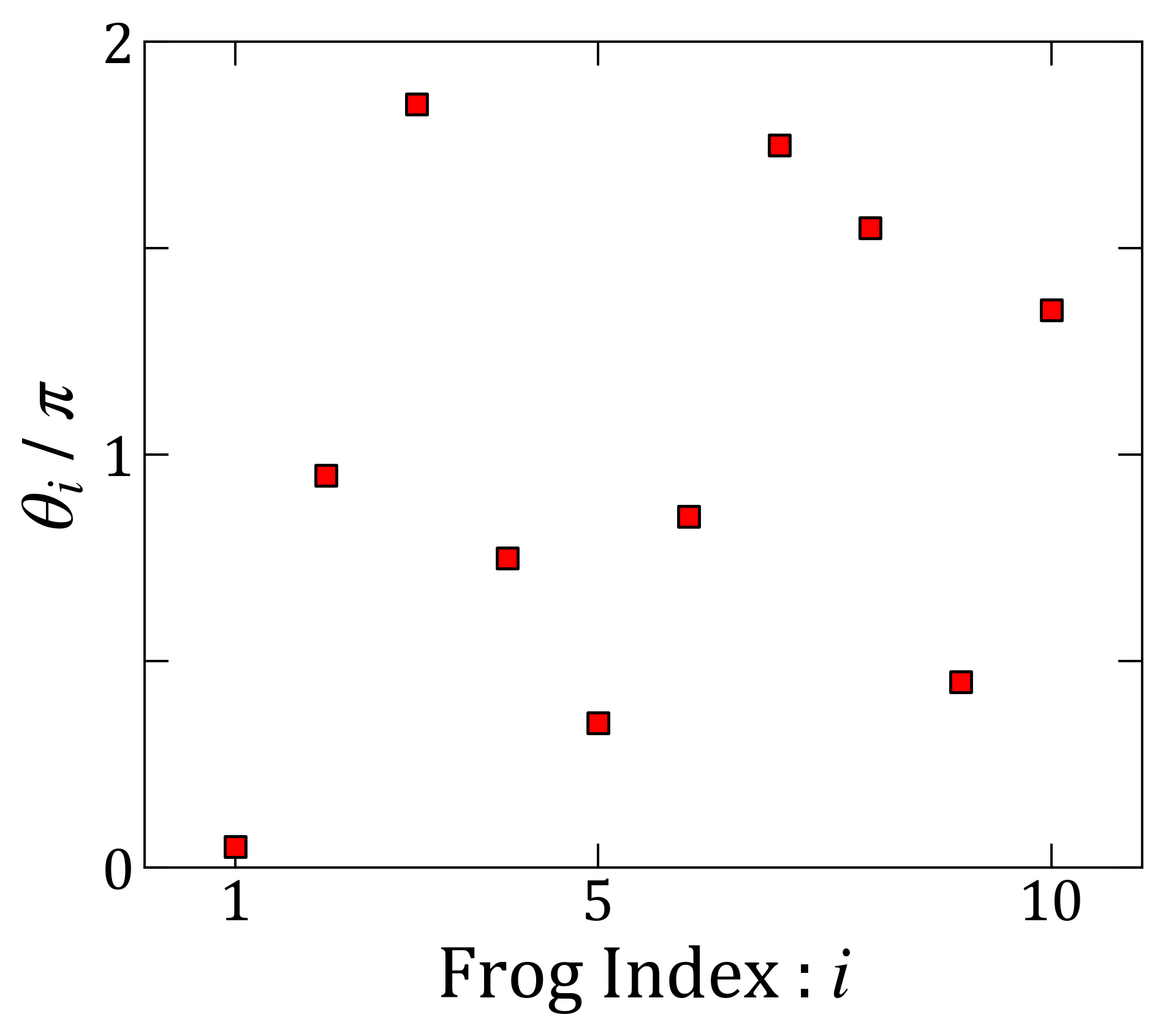}
        \hfill
        \includegraphics[height=3.5cm, keepaspectratio]{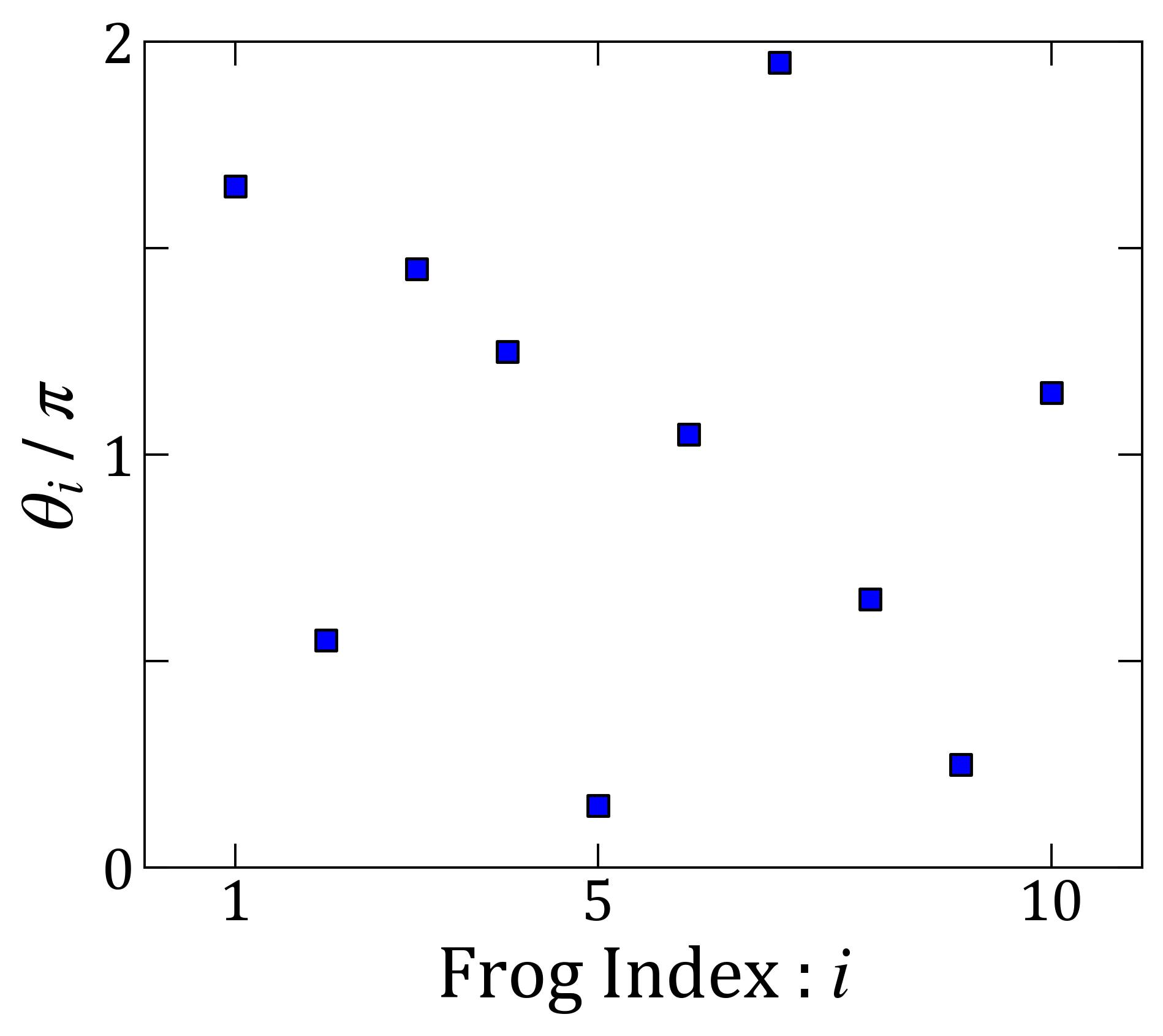}   
        \label{fig:KL0_phaseDist_wavy}        
    \end{subfigure}

    \caption{Wavy antisynchronization obtained from numerical simulation at $\hat{D}_{\mathrm{KL}}=0$. 
    (a) Spatial structure in a frog chorus. The males are positioned along the edge of the field at the same inter-frog distance despite of the species. 
    (b) Wavy antisynchronization in a frog chorus. 
    Here the index $n$ is assigned to all males of the two species. 
    The males synchronize in $0.9\pi$ not only with neighbors of same species, but also with those of a different species.
    (c) Disruption of wavy antisynchronization within the same species. The wavy antisynchronization between neighbors in the same species is partially disturbed by the interspecific interaction.}
    \label{fig:wavy; D_KL=0}
\end{figure}

Figures \ref{fig:Two-cluster; D_KL=0} and \ref{fig:wavy; D_KL=0} show the results of numerical simulations for different initial conditions. 
These results show the existence of two distinct spatio-temporal structures, indicating bistability of the system. 
The first structure (Fig. \ref{fig:Two-cluster; D_KL=0}) demonstrates that (1) all males are positioned along the edge of the field at same inter-frog distances (Fig. \ref{fig:KL0_spatioDist_2cluster}), (2) the males show anti-phase synchronization with neighbors of both the same and different species, forming a two-cluster antisynchronization across the entire chorus (Fig. \ref{fig:KL0_GlobalphaseDist_2cluster}), and (3) some pairs of neighboring males of the same species show in-phase synchronization (Fig. \ref{fig:KL0_phaseDist_2cluster}; e.g., $i=3$ and $4$ for species 1). 
The second structure (Fig. \ref{fig:wavy; D_KL=0}) demonstrates that (1) all males are positioned along the edge of the field, as in the first structure (Fig. \ref{fig:KL0_spatioDist_wavy}), (2) the males show a phase difference of $0.9\pi$ with their neighbors, forming a wavy antisynchronization across the entire chorus (Fig. \ref{fig:KL0_GlobalphaseDist_wavy}), and (3) the wavy antisynchronization between neighbors in the same species is partially disturbed by the interspecific interaction. (Fig. \ref{fig:KL0_phaseDist_wavy}). 
It should be noted that, when the initial condition was changed, wavy antisynchronization with a different phase difference of $1.1\pi$ was also observed. 

\section*{Appendix D: Expected value of $R_s$ at $\hat{D}_{\mathrm{KL}}=0$}

Here, we estimate the expected value of $R_s$ at $\hat{D}_{\mathrm{KL}}=0$ as $\mathrm{E}[R_s(0)]$. 
The purpose of this calculation is to show that the lower bound with the mean value of $R_s \approx 0.3$ in Figure \ref{figs:ckl_Rs_df} can be explained by $\mathrm{E}[R_s(0)]$ on the assumption of the two equilibrium states: two-cluster and wavy antisynchronization.

We have already shown that, at \(\hat{D}_{\mathrm{KL}}=0\), the males form two-cluster and wavy antisynchronization across the entire chorus, depending on the initial condition (see Figs. \ref{fig:Two-cluster; D_KL=0} and \ref{fig:wavy; D_KL=0} in Appendix C).
Using the law of total expectation, the expected value $\mathrm{E}[R_s(0)]$ can be expressed as the weighted sum of the expected values under these two states:
\begin{equation}
\label{eq: E[Rs] in DKL=0, full}
\begin{split}
	\mathrm{E}[R_s(0)] &= \alpha \mathrm{E}[R_s(0)\mid \text{two-cluster}] \\
		&\qquad+ (1-\alpha) \mathrm{E}[R_s(0)\mid \text{wavy}].
\end{split}
\end{equation}
Here, $\mathrm{E}[R_s(0)\mid \text{two-cluster}]$ and $\mathrm{E}[R_s(0)\mid \text{wavy}]$ denote the expected values of $R_s(0)$ at two-cluster synchronization and wavy antisynchronization, respectively. 
The coefficient $\alpha$ is the probability that the two-cluster antisynchronization occurs.
We have estimated the value of $\alpha$ using the result of numerical simulation in Fig. \ref{fig:phi_DKL_ckl30} that represents the distribution of $\theta_{i+1}-\theta_i$ at $\hat{D}_{\mathrm{KL}}=0$. 
It should be noted that, in the Figure, the peaks at $\theta_{i+1}-\theta_i=0$ and $\pi$ correspond to two-cluster antisynchronization. 
Subsequently, the value of $\alpha$ can be estimated from the fraction of the distribution of the two peaks of $\theta_{i+1}-\theta_i=0$ and $\pi$.
According to the above procedure, we have estimated the value of the coefficient as $\alpha \approx 0.57$.

To calculate $\mathrm{E}[R_s(0)]$ according to Equation (\ref{eq: E[Rs] in DKL=0, full}), we need to evaluate each expected value \(\mathrm{E}[R_s(0)\mid \text{two-cluster}]\) and \(\mathrm{E}[R_s(0)\mid \text{wavy}]\).
For that purpose, we first introduce $\mathrm{E}[R_s(\hat{D}_\mathrm{KL})]$ and describe it using the phase difference $\theta_{i+1}-\theta_i$.
Specifically, we use the linearity of expectation and the definition of $R_s$ (Eq. \ref{eq:Rs}) to express \(\mathrm{E}[R_s(\hat{D}_\mathrm{KL})]\) as follows:
\begin{align}
\mathrm{E}[R_s(\hat{D}_\mathrm{KL})] = - \frac{1}{N_s}\sum_{i=1}^{N_s} \mathrm{E}[\cos(\theta_{i+1} - \theta_i;\hat{D}_\mathrm{KL})].
\end{align}
From the simulations of Figures \ref{fig:KL1_spatioDist}, \ref{fig:KL05_spatioDist}, \ref{fig:KL0_spatioDist_2cluster}, and \ref{fig:KL0_spatioDist_wavy}, we assume that all frogs are positioned along the edge of the field but the two species are randomly mixed depending on the initial conditions. 
Under this assumption, the expectation \(\mathrm{E}[\cos(\theta_{i+1} - \theta_i;\hat{D}_\mathrm{KL})]\) can be assumed to be equal for all \(i\).
Therefore, we simplify \(\mathrm{E}[R_s]\) as follows:
\begin{align}
    \mathrm{E}[R_s(\hat{D}_\mathrm{KL})] = - \mathrm{E}[\cos(\theta_{i+1} - \theta_i;\hat{D}_\mathrm{KL})].
\label{eq: general E[R_s]}
\end{align}

Let us evaluate $\mathrm{E}[R_s(0)\mid \text{two-cluster}]$ on the basis of Equation (\ref{eq: general E[R_s]}).
When two-cluster antisynchronization at $\hat{D}_{\mathrm{KL}}=0$ is realized, all pairs of neighboring males show anti-phase synchronization regardless of species (Fig. \ref{fig:KL0_GlobalphaseDist_2cluster}).
Therefore, it is possible that there are several males of the other species between the focal pair of the same species (see Figures [IDs]). 
Consequently, we can describe the phase difference between neighboring males of the same species as follows:
\begin{align}
    \theta_{i+1}-\theta_i=(h+1)\pi.
\end{align}
Here, $h$ ($h=0,\dots,N-N_s$) represents the number of frogs of the different species that exist between the $i$th and $(i+1)$th frogs.
Accordingly, $\mathrm{E}[R_s(0)\mid \text{two-cluster}]$ is affected by the probability $P(h)$ in which a specific value $h$ occurs. 
The point is that, when we fix the species of the $i$th frog, the total number of possible arrangements of frogs is given by the combination $\binom{N-1}{N_s-1}$. 
If exactly $h$ frogs of the different species exist between the $i$th and $(i+1)$th frogs, the number of possible arrangements is given by the combination $\binom{N-h-2}{N_s-2}$.
Consequently, $P(h)$ is described by the following Equation: 
\begin{align}
    P(h) = \frac{\binom{N-h-2}{N_s-2}}{\binom{N-1}{N_s-1}}.
    \label{eq: P(h)}
\end{align}
Based on this result, \(\mathrm{E}[R_s(0)\mid \text{two-cluster}]\) is given by
\begin{align}
    \mathrm{E}[R_s(0)\mid \text{two-cluster}] &= -\mathrm{E}[\cos(\theta_{i+1}-\theta_i) \mid \text{two-cluster}] \notag \\
    &= -\sum_{h=0}^{N-N_s}P(h)\cos((h+1)\pi).
    \label{eq: E[Rs] in 2cluster}
\end{align}

Next, we evaluate $\mathrm{E}[R_s(0)\mid \text{wavy}]$ on the basis of Equation (\ref{eq: general E[R_s]}).
When wavy antisynchronization at $\hat{D}_{\mathrm{KL}}=0$ is realized, neighboring males show a phase difference of either $0.9\pi$ or $1.1\pi$ regardless of species (Fig. \ref{fig:KL0_GlobalphaseDist_wavy} shows the case of $0.9\pi$).
As in the case of two-cluster antisynchronization, the phase difference between neighboring males of the same species depends on the number of frogs of the different species between them.
Therefore, the phase difference $\theta_{i+1}-\theta_i$ is described as follows:
\begin{align}
\theta_{i+1}-\theta_i = 0.9(h+1)\pi ~ \text{or}~ 1.1(h+1)\pi.
\end{align}
Because of the properties of the cosine function, both cases produce the same value for any integer $h$ (i.e., $\cos(0.9(h+1)\pi) = \cos(1.1(h+1)\pi)$).
Therefore, in the same way as Equation (\ref{eq: E[Rs] in 2cluster}), $\mathrm{E}[R_s(0)\mid \text{wavy}]$ is given by
\begin{align}
\mathrm{E}[R_s(0)\mid \text{wavy}] &= -\sum_{h=0}^{N-N_s}P(h)\cos(0.9(h+1)\pi).
\label{eq: E[Rs] in wavy}
\end{align}

Finally, we calculate the expected values for each state.
Substituting $N=20$ and $N_s=10$, which are the values used for the numerical simulation in the main manuscript, into Equations (\ref{eq: P(h)}), (\ref{eq: E[Rs] in 2cluster}), and (\ref{eq: E[Rs] in wavy}) yields
\begin{align}
\mathrm{E}[R_s(0)\mid \text{two-cluster}] &\approx 0.303,\\
\mathrm{E}[R_s(0)\mid \text{wavy}] &\approx 0.300.
\end{align}
Furthermore, we substituted these values into Equation (\ref{eq: E[Rs] in DKL=0, full}) and estimated the expected value of the order parameter as $\mathrm{E}[R_s(0)] \approx 0.302$ that is consistent with numerical simulation in the main manuscript (i.e., $R_s \approx 0.30$ at \(\hat{D}_{\mathrm{KL}}=0\)).

\printbibliography

\end{document}